\documentclass[english,final]{IEEEtran}
\usepackage[T1]{fontenc}
\usepackage[latin9]{inputenc}
\usepackage{verbatim}
\usepackage{float}
\usepackage{bm}
\usepackage{amstext}
\usepackage{amssymb}
\usepackage{graphicx}
\usepackage{esint}

\makeatletter

\newcommand{\lyxmathsym}[1]{\ifmmode\begingroup\def\b@ld{bold}
  \text{\ifx\math@version\b@ld\bfseries\fi#1}\endgroup\else#1\fi}

\floatstyle{ruled}
\newfloat{algorithm}{tbp}{loa}
\providecommand{\algorithmname}{Algorithm}
\floatname{algorithm}{\protect\algorithmname}

  \newtheorem{asm}{Assumption}
  \newenvironment{asmQED}{\begin{asm}}{~\hfill\IEEEQEDclosed\end{asm}}
  \newtheorem{lemQED}{Lemma}
  \newenvironment{lyxLemQED}{\begin{lemQED}}{~\hfill\IEEEQEDclosed\end{lemQED}}
  \newtheorem{defQED}{Definition}
  \newenvironment{lyxDefQED}{\begin{defQED}}{~\hfill\IEEEQEDclosed\end{defQED}}
  \newtheorem{thmQED}{Theorem}
  \newenvironment{lyxThmQED}{\begin{thmQED}}{~\hfill\IEEEQEDclosed\end{thmQED}}
  \newtheorem{corQED}{Corollary}
  \newenvironment{lyxCorQED}{\begin{corQED}}{~\hfill\IEEEQEDclosed\end{corQED}}
  \newtheorem{remrk}{Remark}

\usepackage{cite}

\author{Junting~Chen,~\IEEEmembership{Student~Member,~IEEE}        and~Vincent~K.~N.~Lau,~\IEEEmembership{Fellow,~IEEE}\\

\thanks{ Copyright (c) 2012 IEEE. Personal use of this material is permitted. However, permission to use this material for any other purposes must be obtained from the IEEE by sending a request to pubs-permissions@ieee.org.}%

\thanks{The authors are with the Department of Electronic and Computer Engineering (ECE), The Hong Kong University of Science and Technology         (HKUST), Hong Kong (e-mail: \{eejtchen, eeknlau\}@ust.hk).}%

}

\makeatother

\usepackage{babel}
\begin{document}

\title{Large Deviation Delay Analysis of Queue-Aware  Multi-user MIMO Systems
with Two-timescale Mobile-Driven Feedback}

\maketitle

\begin{abstract}
Multi-user multi-input-multi-output (MU-MIMO) systems transmit data
to multiple users simultaneously using the spatial degrees of freedom
with user feedback channel state information (CSI). Most of the existing
literatures on the reduced feedback user scheduling focus on the throughput
performance and the user queueing delay is usually ignored. As the
delay is very important for real-time applications, a low feedback
queue-aware user scheduling algorithm is desired for the MU-MIMO system.
This paper proposed a two-stage queue-aware user scheduling algorithm,
which consists of a queue-aware mobile-driven feedback filtering stage
and a  user scheduling stage, where the feedback filtering policy
is obtained from an optimization. We evaluate the queueing performance
of the proposed scheduling algorithm by using the sample path large
deviation analysis. We show that the large deviation decay rate for
the proposed algorithm is much larger than that of the CSI-only user
scheduling algorithm. The numerical results also demonstrate that
the proposed algorithm performs much better than the CSI-only algorithm
requiring only a small amount of feedback. \end{abstract}
\maketitle
\IEEEpeerreviewmaketitle

\section{Introduction}

MIMO is an important core technology for next generation wireless
systems. In particular, in multi-user MIMO (MU-MIMO) systems, a base
station (BS) (with $M$ transmit antennas) communicates with multiple
mobile users simultaneously using the spatial degrees of freedom at
the expense of knowledge of channel states at the transmitter (CSIT).
It is shown in \cite{yoo2006optimality,sharif2005capacity} that using
simple zero-forcing precoder and near orthogonal user selection, a
sum rate of $M\log\log K$ can be achieved with full CSIT knowledge
over $K$ users. Yet, full CSIT knowledge is difficult to achieve
in practice and there are a lot of works focusing on reducing the
feedback overhead in MIMO systems \cite{Xia06,Zheng08,bayesteh2008user,zhang2007mimo,sanayei2007opportunistic,diaz2008asymptotic}.
For instance, in \cite{Xia06,Zheng08}, the authors have focused on
the codebook design and performance analysis under limited-rate feedback
schemes. In \cite{bayesteh2008user,zhang2007mimo,sanayei2007opportunistic},
on the other hand, a threshold based feedback control is adopted where
users attempt to feedback only when its channel quality exceeds a
threshold. It was further shown that a sum rate capacity $\mathcal{O}(M\log\log K)$
can be achieved when only $\mathcal{O}(M\log\log\log K)$ users feeding
back to the BS \cite{bayesteh2008user}. 

While there are a lot of works that consider reduced feedback design
for MU-MIMO, all these existing works focused on the throughput performance.
They have assumed infinite backlog at the base station and therefore,
ignored the bursty arrival of the data source as well as the associated
delay performance, which is very important for real-time applications.
For instance, the CSI information indicates \emph{good opportunity
to transmit} whereas the \emph{Queue State Information} (QSI) indicates
the \emph{urgency} of the data flow. A delay-aware MU-MIMO system
should incorporate both the CSI and QSI in the user scheduling. However,
it is far from trivial to integrate these information in determining
the user priority. There are some works considering QSI in the user
scheduling of MU-MIMO systems. In \cite{Cui11}, the author considered
a queue-aware power control and dynamic clustering in downlink MIMO
systems. In \cite{she2009joint}, the authors considered MU-MIMO user
scheduling to maximize queue-weighted sum rate. Due to the exponentially
large solution space, heuristic greedy-based algorithm is proposed.
However, these works required the BS to have global CSI knowledge
of all the users, which is hard to achieve in practice. Furthermore,
the delay performance in \cite{she2009joint} is obtained by simulation
only and not much design insights can be obtained in these works.
In general, there are still a number of first order technical challenges
associated with designing delay-aware MU-MIMO systems. 
\begin{itemize}
\item \textbf{Challenges in User Scheduling Design}: For real-time applications,
it is important to exploit CSI and QSI in the user scheduling. Yet,
it is highly non-trivial to design a \emph{priority metric} that strike
a balance between transmission opportunity and urgency. One one hand,
the Markov decision process (MDP) based methods \cite{Djonin07,Fu10}
result in high complexity (exponential w.r.t. $K$).  On the other,
brute-force application of Lyapunov optimization techniques \cite{neely2005dynamic}
in MU-MIMO is also not feasible because of the associated exponential
complexity of user selection for MU-MIMO.
\item \textbf{Challenges in Delay Analysis}: Due to the QSI-aware control
algorithm, the service rate of the data queues are \emph{state-dependent}
and the queue dynamics from these $K$ data flows are coupled together.
This makes the queueing delay analysis extremely difficult. There
is no closed form results on the steady state distributions of the
queue length in such complex queueing systems. In \cite{huang2009stability},
the authors characterized the \emph{stability region} of the MU-MIMO
systems under limited CSI feedback. Yet, stability is only a weak
form of delay performance. 
\end{itemize}

In this paper, we consider a MU-MIMO downlink system with a $M$-antenna
BS and $K$ multi-antenna mobile users. The BS applies the \emph{random
beamforming} for MU-MIMO to exploit the multi-user diversity. To overcome
the complexity challenge of user scheduling, we propose  a two-timescale
delay-aware user scheduling policy for the MU-MIMO system. The proposed
policy consists of two stages, namely the \emph{queue-aware user-driven
feedback filtering stage} and the \emph{dynamic queue-weighted user
scheduling stage}. At the first stage (slower timescale), the BS broadcasts
a QSI-dependent user feedback candidate list and only the mobiles
in the list are allowed to feedback the CSI to the BS. At the second
stage (faster timescale), the BS selects the best user according to
the queue-weighted metric among the users selected in the first stage.
Based on the two-timescale user scheduling policy, we then analyze
the delay performance of the MU-MIMO system. It is in general difficult
to analyze the delay for state-dependent coupled queues. To overcome
this challenge, we consider the large deviation tail for the maximum
queue length among all the users, which reflects the worse case delay
performance in the system. Using large deviation theory for random
process \cite{shwartz1995large}, we derive the asymptotic exponential
decay rate for the tail probability of the maximum queue length. Specifically,
we quantify the asymptotic decay rate $-\frac{1}{B}\log(\mbox{Pr}(\max_{k}Q_{k})>B)$
as buffer size $B\to\infty$. We show that the decay rate of the worst
case queue length of  the proposed delay-aware scheduling algorithm
scales as $\mathcal{O}(\log K)$, which is substantially better than
traditional MU-MIMO user scheduling baseline schemes. 

The rest of the paper is organized as follows. We present the system
model, bursty data source and queueing model and the proposed two-timescale
delay-aware user scheduling policy in Section \ref{sec:System-Model}.
In Section \ref{sec:Delay-Aware-control}, we derive the optimal user-driven
feedback filtering strategy using Lyapunov approach. We then analyze
the maximum queue length property using sample path fluid approximation
and large deviation theory in Section \ref{sec:Queueing-delay-analysis}.
Numerical results are provided in Section \ref{sec:Numerical-Results}
and we conclude the results in Section \ref{sec:Conclusions}.

\section{System Model\label{sec:System-Model}}

\subsection{MU-MIMO System Model}

We consider a downlink MU-MIMO system with a $M$-antenna BS and $K$
geometrically dispersed mobile users ($K\gg M$). Each mobile user
has $N$ receive antennas. Using MU-MIMO techniques, the BS transmits
$M$ data streams to a group of selected users at each time slot.
The wireless channel between each user and the BS is modeled as a
Rayleigh fading channel. Specifically, the received signal $\mathbf{y}_{k}\in\mathbb{C}^{N\times1}$
by the user $k$ is given by 
\begin{equation}
\mathbf{y}_{k}=\sqrt{P}H_{k}\mathbf{x}+\mathbf{n}_{k}\qquad\forall k\in\mathcal{A}(t)\label{eq:channel-model}
\end{equation}
where $ $$\mathbf{x}\in\mathbb{C}^{M\times1}$ is the normalized
transmitted signal with $\mathbb{E}\left[\mbox{Tr}(\mathbf{xx}^{*})\right]=M$,
i.e., the normalized transmit power on each antenna is assumed to
be one, $H_{k}\in\mathbb{C}^{N\times M}$ is the zero mean, unit-variance
circularly symmetric complex Gaussian channel matrix from the transmitter
to the user $k$, $\mathbf{n}_{k}\in\mathbb{C}^{N\times1}\sim\mathcal{CN}(\mathbf{0},\mathbf{I}_{N})$
is the Gaussian additive noise vector, $P$ is the transmit power
at the BS, and $\mathcal{A}(t)$ denotes the set of the scheduled
users at time slot $t$. We have the following assumption on the channel
matrices $\{H_{k}\}$. 
\begin{asmQED}
[Assumptions on Channel Matrices]\label{asm:channel-model} The
channel matrix $H_{k}(t)$ is a $N\times M$ complex matrix for user
$k$, where each element $h_{k}^{(i,j)}(t)$ has a zero mean unit
variance stationary Gaussian distribution $\mathcal{CN}(0,1)$, and
autocorrelation function $\mathcal{R}_{k}^{(i,j)}(\tau)$. It is assumed
that $\mathcal{R}_{k}^{(i,j)}(\tau)\to0$, exponentially fast as $\tau\to\infty$.
The mobile users are assumed to have perfect knowledge of their local
CSI. However, only a selected subset of users will feedback their
CSI to the BS and the feedback information is delivered through a
noiseless feedback channel. 
\end{asmQED}

The above channel assumptions have captured many practical channel
models, such as the i.i.d. model and the AR($n$) model \cite{Baddour01}.

At the BS, random beamforming is used to support near-orthogonal data
streams transmissions to the selected users without knowing the full
CSI%
\footnote{Note that the proposed two-timescale framework can also work for other
beamforming schemes, such as zero-forcing. One may derive the corresponding
control policy using similar techniques presented in this paper.%
}. The BS chooses $M$ random orthonormal vectors $\{\phi_{1},\dots,\phi_{M}\}$,
where $\phi_{m}\in\mathbb{C}^{M\times1}$ are generated according
to an isotropic distribution. Let $\mathbf{s}(t)=(s_{1}(t),\dots,s_{M}(t))$
be the vector of the transmit symbols. The transmit signal is given
by 
\[
\mathbf{x}(t)=\sum_{m=1}^{M}\phi_{m}s_{m}(t).
\]
Therefore, the receive signal at the $k$-th user is  
\[
\mathbf{y}_{k}(t)=\sum_{m=1}^{M}\sqrt{P}H_{k}\phi_{m}s_{m}(t)+\mathbf{n}_{k}.
\]

We assume the receivers know the beamforming vectors $\{\phi_{m}\}$.
The \emph{effective SINR} of the $i$-th beam on the $n$-th receive
antenna of the $k$-th user can be calculated as follows,  
\begin{equation}
\mbox{SINR}_{k,n}^{i}=\frac{\left|H_{k}^{(n)}\phi_{i}\right|^{2}}{\sum_{j,j\neq i}\left|H_{k}^{(n)}\phi_{j}\right|^{2}+1/P}.\label{eq:effective-SINR}
\end{equation}
where $H_{k}^{(n)}$ denotes the $n$-th row of the channel matrix
$H_{k}$ of user $k$.  By selecting the users with the highest
SINR on each beam, the transmitter can support near-orthogonal transmissions
and exploit multi-user diversity  without the global CSI $\{H_{k}\}$
\cite{chung2003random}.

\subsection{Bursty Data Source and Queue Model}

Data arrives in packets randomly for different users. Let $A_{k}(t)$
denote the number of packets that arrive at the BS for user $k$ during
time slot $t$, and $\mathbf{A}(t)=(A_{1}(t),\dots,A_{K}(t))$. We
assume that the arrivals $A_{k}(t)$ are i.i.d over different time
slot $t$. We have the following assumptions regarding the bursty
arrival processes $A_{k}(t)$.
\begin{asmQED}
[Bursty Source Model]\label{asm:source-model} The packet arrival
$A_{k}(t)$ are identically and independently distributed (i.i.d.)
with respect to (w.r.t.) $t$ and independent w.r.t. $k$ according
to a general distribution with mean $\mathbb{E}[A_{k}(t)]=\lambda_{k}$
and finite moment generating function (MGF) $\psi_{A,k}(\theta)=\mathbb{E}\left[e^{\theta A_{k}}\right]$.
The packet length is assumed to be constant $L$ bits.  
\end{asmQED}

The BS maintains queueing backlogs $Q_{k}(t)$ for each user $k$.
Let $D_{k}(\mathbf{Q}(t),\mathbf{H}(t))$ represents the amount of
departure in packets for user $k$ at time slot $t$, where $\mathbf{Q}(t)=(Q_{1}(t),\dots,Q_{K}(t))$
and $\mathbf{H}(t)=(H_{1}(t),\dots,H_{K}(t))$. $D_{k}(\centerdot)$
depends on the specific user scheduling policy. The queueing dynamics
for user $k$ is given by 
\begin{equation}
Q_{k}(t+1)=\left[Q_{k}(t)-D_{k}(\mathbf{Q}(t),\mathbf{H}(t))\right]^{+}+A_{k}(t)\label{eq:queue-dynamic}
\end{equation}
where the operator $[\centerdot]^{+}$ represents $[w]^{+}=\max\{0,w\}$.
Here we do not consider packet drops or retransmissions. Using Little\textquoteright{}s
Law \cite{Little1961:law}, the average delay of the $k$-th user
is given by $\overline{T}_{k}=\overline{Q}_{k}/\overline{D}_{k}$,
where $\overline{Q}_{k}$ is the average backlog for the $k$-th queue
and $\overline{D}_{k}$ is the average departure at each time slot.
As a result, there is no loss of generality to study the queue length
$Q_{k}$ for the purpose of understanding the delay. Obviously, the
queue length (or the delay) of the MU-MIMO system depends on how we
use the channel resources. Hence the goal of the user scheduling controller
is to adjust the channel access opportunity for all the users so that
their queue lengths (or delay) are minimized while maintaining a high
system throughput.

\subsection{Two-timescale User Scheduling with Reduced Feedback for MU-MIMO\label{sub:Two-Timescale-scheduling}}

A reasonable delay-aware user scheduling algorithm should jointly
adapt to both the CSI (to capture good transmission opportunity) and
the QSI (to capture the urgency). In particular, we are interested
in the control policy that can maximize queue stability region. However,
conventional throughput optimal (in stability sense) user scheduling
policies such as max-weighted-queue (MWQ) algorithms \cite{neely2005dynamic}
require global CSI and QSI knowledge.  However, the CSI is available
at the mobile user side while the QSI is available at the BS. Furthermore,
the MWQ policy requires solving a queue weighted sum rate combinatorial
optimization problem, which has exponential searching space. Hence,
a brute-force solution of the MWQ problem requires huge signaling
overhead as well as huge complexity. To overcome these challenges,
we propose a two-timescale user scheduling solution as follows. 
\begin{itemize}
\item Stage I: \emph{Queue-aware user-driven feedback filtering}. The BS
determines and broadcasts the user feedback probability $\{p_{1}(\mathbf{Q}),\dots p_{K}(\mathbf{Q})\}$
based on the user queueing backlogs $\mathbf{Q}(t)$ for every $T$
time slots. Mobile user $k$ randomly feedback to the BS in the stage
II with probability $p_{k}$. We denote $\chi_{k}\in\{0,1\}$ as the
stochastic feedback filtering policy with $\mbox{P}(\chi_{k}=1)=p_{k}$,
and a user $k$ feeds back when $\chi_{k}(t)=1$. The motivation of
the mobile feedback filtering is to save the feedback cost by reducing
the lower priority users from feeding back. 
\item Stage II: \emph{Dynamic Queue-Weighted User Scheduling}. If the feedback
indicator $\chi_{k}=1$, then user $k$ measures the effective SINR
vector $\{\mbox{SINR}_{k,n}^{1},\lyxmathsym{\ldots},\mbox{SINR}_{k,n}^{M}\}$
on each receive antenna $n$ according to (\ref{eq:effective-SINR})
and finds the strongest beam $i^{*}(k,n)=\arg\max_{1\leq i\leq M}\mbox{SINR}_{k,n}^{i}$.
The mobile then feeds back the selected beam index $i^{*}(k,n)$ and
the associated $\mbox{SINR}_{k,n}^{i^{*}(k,n)}$ to the BS on each
$n$. The set of feedback users at time slot $t$ is denoted by $\mathcal{F}(t)$.
The BS schedules user $k^{*}(i)$ to transmit at the $i$-th beam
to maximize the queue-weighted throughput, i.e., $k^{*}(i)=\arg\max_{k\in\mathcal{F}(t)}Q_{k}\log\left(1+\gamma_{k}^{i}\right)$,
where $\gamma_{k}^{i}=\max_{n\in\mathcal{N}(k,i)}\mbox{SINR}_{k,n}^{i}$
denotes the highest SINR of user $k$ on the $i$-th beam%
\footnote{We define $\gamma_{k}^{i}=0$ if $\mathcal{N}(k,i)=\emptyset$.%
} over $n\in\mathcal{N}(k,i)$. Here $\mathcal{N}(k,i)=\left\{ n:1\leq n\leq N,i^{*}(k,n)=i\right\} $
denotes the set of receive antennas of user $k$ that have fed back
the SINR for the $i$-th beam%
\footnote{Although we have assumed the fading channels are i.i.d. among users,
the two-timescale algorithm framework can also be applied to non-i.i.d.
users, using a similar feedback policy in stage I. However, the analysis
in this case is much more complicated, and we shall leave it to the
future work.%
}. As a result, the stage II user scheduling exploits the multi-user
diversity among the set of users attempting to feedback $\mathcal{F}(t)$. 
\end{itemize}

The following lemma shows that, in a MU-MIMO system, it is sufficient
for each user feeding back only the beam with the highest SINR as
Stage II policy suggests.
\begin{lyxLemQED}
[SINR property of a MU-MIMO channel \cite{sharif2005capacity}]\label{lem:tightness-single-beam-feedback}
If $\max_{k\in\mathcal{F},1\leq n\leq N}\mbox{SINR}_{k,n}^{i}\geq1$,
$\forall i=1\dots M$, then it is impossible for a user to have maximum
SINRs for more than two beams on one antenna, i.e., for $(k^{*},n^{*})=\arg\max_{k\in\mathcal{F},1\leq n\leq N}\mbox{SINR}_{k,n}^{i}$,
we have $\mbox{SINR}_{k^{*},n^{*}}^{i}=\max_{1\leq j\leq M}\mbox{SINR}_{k^{*},n^{*}}^{j}$,
$\forall i$. 
\end{lyxLemQED}

One may easily see that the probability for violating the condition
in Lemma \ref{lem:tightness-single-beam-feedback} exponentially decreases
w.r.t. the number of feedback users, and hence is negligible. 

\begin{figure}
\begin{centering}
\includegraphics[width=1\columnwidth]{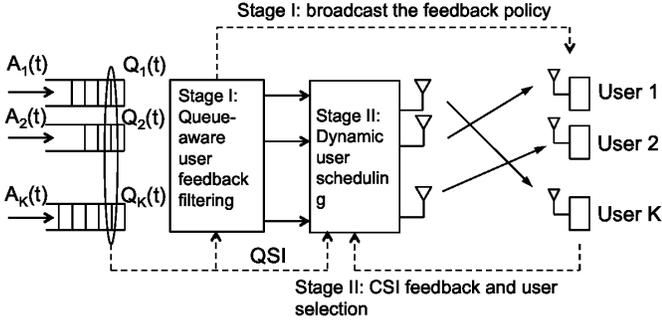}
\par\end{centering}

\caption{\label{fig:two-timescale}The two stage joint CSI and QSI user scheduling
in a multi-user MIMO system. At stage I, the BS determines the user
feedback priority based on the QSI. At stage II, a portion of selected
users feedback their CSI and the BS schedules users for transmission
based on their CSI feedback.}
\end{figure}

Fig. \ref{fig:two-timescale} depicts an illustration of the two stages
user scheduling policy. The policy tries to balance the transmission
opportunity and urgency with a low complexity and low feedback cost
strategy. For the user with a long queue, it will be given priority
to feedback during the stage I feedback filtering phase. Users who
have passed the stage I filtering will compete for channel access
based on the stage II queue weighted scheduling in which users with
better queue weighted metric will be served. Moreover, the two stages
processing can be implemented on different timescales. The SINR feedback
and user scheduling in stage II is done at every time slot $t$, while
the user feedback probability $\{p_{k}(\mathbf{Q})\}$ determined
in stage I can be updated once every $T$ time slots. The update period
$T$ trades the performance of the two-timescale policy with the control
signaling overhead. With a larger $T$, there is a smaller signaling
overhead associated with broadcasting $\{p_{k}(\mathbf{Q})\}$ in
stage I but then the feedback priority may be driven by outdated QSI. 

\subsection{Queue-Aware Feedback Filtering (Stage I) Optimization}

The feedback filtering control in stage I plays a critical role in
the overall delay performance of the MU-MIMO system. In the following,
we adopt a Lyapunov optimization technique to derive the stage I feedback
filtering policy to achieve the maximum \emph{queue stability region}
in the MU-MIMO system.

\subsubsection{Queue Stability }

We first define the queue stability and the stability region formally
below. 
\begin{lyxDefQED}
[Queue Stability]\label{def:Queue-Statiblity} The queueing system
is called \emph{stable }if $\lim\sup_{t\to\infty}\frac{1}{t}\mathbb{E}\left[\max_{k}Q_{k}(t)\right]<\infty$.
\end{lyxDefQED}
\begin{lyxDefQED}
[Stability region and Throughput Optimal]\label{def:Stability-region}
The \emph{stability region} $\mathcal{C}$ is the closure of the set
of all the arrival rate vectors $\{\lambda_{k}\}$ that can be stabilized
in a MU-MIMO system for some feedback probability vector $\{p_{k}\}$
in the two-timescale scheduling framework. A \emph{throughput optimal}
\emph{feedback control} is a feedback probability vector $\{p_{k}\}$
that  stabilizes all the arrival rate vectors $\{\lambda_{k}\}$ within
the stability region $\mathcal{C}$. 
\end{lyxDefQED}

\subsubsection{The Data Rate and the Amount of Feedback}

Let $J_{k}^{i}(\mathbf{Q},\mathbf{H},\bm{\chi})\in\{0,1\}$ be the
scheduling indicator of the $k$-th user on the $i$-th beam according
to the Stage II policy. Therefore, the instantaneous data rate for
user $k$ is given by 
\begin{equation}
R_{k}(\mathbf{Q},\mathbf{H},\bm{\chi})=\sum_{i=1}^{M}J_{k}^{i}(\mathbf{Q},\mathbf{H},\bm{\chi})\chi_{k}\log(1+\gamma_{k}^{i}).\label{eq:data-rate}
\end{equation}

We define the conditional feedback cost $\mathcal{S}(\mathbf{Q})$
and the average feedback cost $\overline{\mathcal{S}}$ as follows,
\begin{equation}
\mathcal{S}(\mathbf{Q})=\mathbb{E}\left[\sum_{k}\chi_{k}|\mathbf{Q}\right]=\sum_{k}p_{k}(\mathbf{Q}),\quad\mbox{and}\quad\overline{\mathcal{S}}=\mathbb{E}\left[\mathcal{S}(\mathbf{Q})\right].\label{eq:feedback-cost}
\end{equation}
In addition, the minimum average feedback cost to achieve the maximum
queue stability region $\mathcal{C}$ in the MU-MIMO system is denoted
as $\overline{\mathcal{S}}^{*}$.

\subsubsection{The Feedback Filtering Optimization}

The feedback filtering control policy is derived from the Lyapunov
technique and to achieve the throughput optimality.

Define $L(\mathbf{Q})=\sum_{k}Q_{k}^{2}$ as the Lyapunov function.
Then the one-step conditional Lyapunov drift $\triangle L(\mathbf{Q}(t))$
is given by,
\begin{eqnarray}
\triangle L(\mathbf{Q}(t)) & \triangleq & \mathbb{E}\left[L(\mathbf{Q}(t+1)-L(\mathbf{Q}(t))|\mathbf{Q}(t)\right].\label{eq:lyapunov-drift}
\end{eqnarray}
The following lemma establishes the relationship between the Lyapunov
drift (\ref{eq:lyapunov-drift}) and the queue stability.
\begin{lyxLemQED}
[Lyapunov drift and the queue stability]\label{lem:Lyapunov-drift-stability}
Given positive constants $V$ and $\epsilon$, the $K$ queues of
the MU-MIMO system $\{Q_{1}(t),\lyxmathsym{\ldots},Q_{K}(t)\}$ are
stable if the following condition is satisfied,
\begin{equation}
\triangle L(\mathbf{Q}(t))+V\mathbb{E}\left\{ \mathcal{S}(\mathbf{Q}(t))|\mathbf{Q}(t)\right\} \leq C_{0}K-\epsilon\sum_{k}Q_{k}(t)+V\overline{\mathcal{S}}^{*}\label{eq:lyapunov-drift-with-feedback-cost}
\end{equation}
for some constant $C_{0}<\infty$ and all $\mathbf{Q}(t)$. The average
queue length satisfies
\begin{equation}
\sum_{k}\overline{Q}_{k}\triangleq\lim\sup_{T\to\infty}\frac{1}{T}\sum_{\tau=0}^{T-1}\sum_{k}\mathbb{E}\left[Q_{k}(\tau)\right]\leq\frac{C_{0}K+V\overline{\mathcal{S}}^{*}}{\epsilon}\label{eq:average-queue-length}
\end{equation}
and the average feedback cost satisfies 
\begin{equation}
\overline{\mathcal{S}}\triangleq\lim\sup_{T\to\infty}\frac{1}{T}\sum_{\tau=0}^{T-1}\mathcal{S}(\mathbf{Q}(\tau))\leq\overline{\mathcal{S}}^{*}+C_{0}K/V.\label{eq:average-feedback-cost}
\end{equation}

\end{lyxLemQED}
\begin{proof}
The proof can be extended from \cite[Lemma 1]{neely2006energy} by
replacing the power cost function with the feedback cost function
$\mathcal{S}(\mathbf{Q})$ defined in (\ref{eq:feedback-cost}).
\end{proof}

Lemma \ref{lem:Lyapunov-drift-stability} motivates us to minimize
the Lyapunov drift in (\ref{eq:lyapunov-drift-with-feedback-cost})
to achieve the maximum queue stability region. With this insight,
we have the feedback filtering control problem as follows.

\emph{Feedback Filtering Control Problem (FFCP):} Observing the current
queue length $\mathbf{Q}(t)$, users feedback their CSI according
to the probability vector $\mathbf{p}^{*}(\mathbf{Q}(t))=\{p_{1}^{*}(\mathbf{Q}(t)),\dots,p_{K}^{*}(\mathbf{Q}(t))\}$,
where $\mathbf{p}^{*}(\mathbf{Q}(t))$ is obtained from the solution
of the following optimization problem, 
\begin{equation}
\max_{\{0\leq p_{k}\leq1\}}\quad\mathbb{E}\bigg[\sum_{k=1}^{K}Q_{k}(t)R_{k}(\mathbf{Q},\mathbf{H},\bm{\chi})-V\mathcal{S}(\mathbf{Q}(t))\bigg].\label{eq:queue-weighted-sum-rate}
\end{equation}

The parameter $V$ in (\ref{eq:queue-weighted-sum-rate}) trades off
the average queue length (delay) and the feedback cost. A large parameter
$V$ reduces the average feedback cost in (\ref{eq:average-feedback-cost})
but results in a larger average queue length (\ref{eq:average-queue-length}).
Note that due to the feedback filtering variable $\bm{\chi}\in\{0,1\}^{K}$,
we have an exponential complexity (w.r.t. $K$) to evaluate the expectation
in (\ref{eq:queue-weighted-sum-rate}). This makes the problem difficult
to solve. In the next section, we try to derive the solution of the
FFCP problem by exploiting the specific problem structure.

\section{The Queue-Aware User Feedback Filtering  Algorithm\label{sec:Delay-Aware-control}}

In this section, we focus on deriving the FFCP solution to (\ref{eq:queue-weighted-sum-rate}).
Towards this end, we first decompose FFCP into two-level subproblems
and study their properties. We then proceed to find the optimal solution
to the inner problem and derive a low complexity algorithm to find
an approximate solution to the outer problem.

\subsection{Property of the FFCP problem}

Using primal decomposition techniques, (\ref{eq:queue-weighted-sum-rate})
can be transformed into the following two subproblems
\begin{itemize}
\item Inner subproblem:
\begin{eqnarray}
\mathcal{W}(S)=\max_{\{p_{k}\}} & \mathbb{E}\bigg[\sum_{k=1}^{K}Q_{k}(t)R_{k}(\mathbf{Q},\mathbf{H},\bm{\chi})\bigg]\label{eq:inner-subproblem}\\
\mbox{subject to} & 0\leq p_{k}\leq1,\qquad\forall k=1,\dots,K\label{eq:inner-subproblem-c1}\\
 & \sum_{k=1}^{K}p_{k}=S\label{eq:inner-subproblem-c2}
\end{eqnarray}
where $S$ is an auxiliary variable with the meaning of the average
feedback cost (the number of feedback users).
\item Outer subproblem: 
\begin{equation}
\max_{S}\quad\mathcal{W}(S)-VS.\label{eq:outer-subproblem}
\end{equation}

\end{itemize}

The objective function (\ref{eq:inner-subproblem}) of the inner problem
can be written as 
\[
\mathbb{E}\bigg\{\mathbb{E}\bigg[\sum_{k=1}^{K}Q_{k}(t)R_{k}(\mathbf{Q},\mathbf{H},\bm{\chi})\big|\bm{\chi}\bigg]\bigg\}=\sum_{j=1}^{2^{K}}w_{j}(\mathbf{Q})\mbox{Pr}(\bm{\chi}=\bm{\chi}^{(j)})
\]
where $w_{j}(\mathbf{Q})=\mathbb{E}_{\mathbf{H}}\big[\sum_{k=1}^{K}Q_{k}(t)R_{k}(\mathbf{Q},\mathbf{H},\bm{\chi})\big|\bm{\chi}^{(j)}\big]$
is a deterministic parameter independent of $\{p_{k}\}$, and $\mbox{Pr}(\bm{\chi}=\bm{\chi}^{(j)})=\prod_{k}p_{k}^{\chi_{k}^{(j)}}(1-p_{k})^{\chi_{k}^{(j)}}$
is the probability of a particular feedback indicator vector $\bm{\chi}^{(j)}$,
$j=1,\dots,2^{K}$.

The above expression is a posynomial w.r.t. $\{p_{k}\}$. Moreover,
the constraints (\ref{eq:inner-subproblem-c1})-(\ref{eq:inner-subproblem-c2})
are monomials. Therefore, the inner problem is a geometric programming
(GP) \cite{Chiang:2005kx}. A nice property of a GP is that a local
optimum is also a global optimum. However, it is almost impossible
to solve (\ref{eq:inner-subproblem}) following the standard GP techniques,
as it contains $2^{K}$ terms and the closed form expressions $w_{j}(\mathbf{Q})$
may not be available either. In the following, we find an optimal
solution of the inner problem by exploiting the specific structure.

\subsection{Solution to the inner problem\label{sub:data-rate}}

Let $\Pi=\{\pi(1),\dots,\pi(K)\}$ be a permutation of $\mathbf{Q}$
such that $Q_{\pi(1)}\geq Q_{\pi(2)}\geq\dots\geq Q_{\pi(K)}$. We
find the optimal solution of the inner problem under the average feedback
amount $\mathbb{E}\left[\sum\chi_{k}\right]=\sum_{k}p_{k}=S$ as follows.
\begin{lyxThmQED}
[The optimal solution to the inner problem]\label{thm:solution_pk}
The feedback probability $\{p_{k}\}$ to solve (\ref{eq:inner-subproblem})
is given by 
\begin{eqnarray}
p_{\pi(k)} & = & 1,\quad\qquad\;1\leq k\leq\left\lfloor S\right\rfloor \label{eq:optimal-pk}\\
p_{\pi(k_{0})} & = & S-\left\lfloor S\right\rfloor ,\;\; k_{0}=\left\lfloor S\right\rfloor +1\label{eq:optimal-pk-2}\\
p_{\pi(k)} & = & 0,\quad\qquad\;\mbox{otherwise.}\label{eq:optimal-pk-3}
\end{eqnarray}

\end{lyxThmQED}
\begin{proof}
Please refer to Appendix \ref{app:Poof-thm-solution_pk} for the proof.
\end{proof}

Although an intuition may argue that it might be better to allow more
than $S$ users to feed back (each with lower $p_{k}$) in order to
boost up the opportunistic utility in stage II, the above result shows
that the best strategy is actually allowing only the users with the
$S$ largest queues to feed back, while keeping the others inactive.

\subsection{Solution to the outer subproblem}

To derive the optimal feedback cost $S^{*}$, we first study the mean
data rate $\mathbb{E}[R_{k}(\mathbf{Q},\mathbf{H},\bm{\chi})]$ (denoted
as $\overline{R}_{k}$) in the utility function (\ref{eq:inner-subproblem}).
Define $\eta_{k}(S)\triangleq\mathbb{E}\left[R_{k}(\mathbf{Q},\mathbf{H},\bm{\chi})\big|\chi_{k}=1,\sum_{k}\chi_{k}=S\right]$
as the average data rate for user $k$, conditioned on the feedback
amount being $|\mathcal{F}|=S$. We characterize $\eta_{k}(S)$ in
the following lemma.
\begin{lyxLemQED}
[Data rate under heavy traffic approximation]\label{lem:data-rate-strongest-user}
Given the set of feedback users $\mathcal{F}$, where $|\mathcal{F}|=S$.
If $\frac{Q_{\pi(1)}}{Q_{\pi(S)}}\approx1$, then we have for $k\in\mathcal{F}$,
\begin{eqnarray}
\eta_{k}(S)\approx M\int_{0}^{\infty}\log(1+x)Nf(x)F(x)^{NS-1}dx\triangleq\hat{\eta}(S)\label{eq:data-rate-max-user}
\end{eqnarray}
where 
\begin{equation}
F(x)=1-\frac{e^{-x/P}}{(1+x)^{M-1}}.\label{eq:snr_cdf}
\end{equation}
is the cumulative distribution function (CDF) of $\mbox{SINR}_{k,n}^{i}$
in (\ref{eq:effective-SINR}) and $f(x)$ is the corresponding probability
distribution function (PDF).

\end{lyxLemQED}
\begin{proof}
Please refer to Appendix \ref{app:maximum-data-rate} for the proof.
\end{proof}

The approximation is accurate when the ratio $ $$\frac{Q_{\pi(1)}}{Q_{\pi(S)}}$
is close to $1$, which means all the feedback users have comparable
queue lengths. This can usually happen in heavy traffic scenario where
most of the users have large queues. As such, we have 
\begin{eqnarray}
\mathcal{W}(S) & = & \mathbb{E}\left[\sum_{k=1}^{\lfloor S\rfloor}Q_{\pi(k)}R_{\pi(k)}|\chi_{\pi(k_{0})}=0\right]\left(1-p_{\pi(k_{0})}\right)\nonumber \\
 &  & \quad+\mathbb{E}\left[\sum_{k=1}^{\lfloor S\rfloor+1}Q_{\pi(k)}R_{\pi(k)}|\chi_{\pi(k_{0})}=1\right]p_{\pi(k_{0})}\nonumber \\
 & \approx & \sum_{k=1}^{\lfloor S\rfloor}Q_{\pi(k)}\hat{\eta}(\lfloor S\rfloor)\left[1-\left(S-\lfloor S\rfloor\right)\right]\label{eq:utility-func-W(s)}\\
 &  & \quad+\sum_{k=1}^{\lfloor S\rfloor+1}Q_{\pi(k)}\hat{\eta}(\lfloor S\rfloor+1)\left(S-\lfloor S\rfloor\right)\triangleq\hat{\mathcal{W}}(S).\nonumber 
\end{eqnarray}
and we obtain an approximation to the outer problem (\ref{eq:outer-subproblem})
as 
\begin{eqnarray}
\max_{S\leq K} & \hat{\mathcal{U}}(S)\triangleq\hat{\mathcal{W}}(S)-VS.\label{eq:outer-subproblem-2}
\end{eqnarray}

Problem (\ref{eq:outer-subproblem-2}) is concave and has a nice property
as shown in the following.
\begin{lyxThmQED}
[Solution property of (\ref{eq:outer-subproblem-2})]\label{thm:outer-concavity-integer}
The objective function $\hat{\mathcal{U}}(S)$ in (\ref{eq:outer-subproblem-2})
is concave. Moreover, the optimal solution $S^{*}$ is an integer.
\end{lyxThmQED}
\begin{proof}
Please refer to Appendix \ref{app:proof-thm-concave-integer} for
the proof.
\end{proof}

Theorem \ref{thm:outer-concavity-integer} suggests that a bisection
algorithm can be applied to find the unique solution $S^{*}$ in (\ref{eq:outer-subproblem-2})
in at most $\log_{2}(K)$ steps, where the optimality condition can
be expressed as 

\begin{equation}
\hat{\mathcal{U}}(S^{*})\geq\hat{\mathcal{U}}(S^{*}+1)\mbox{ and }\hat{\mathcal{U}}(S^{*})\geq\hat{\mathcal{U}}(S^{*}-1)\label{eq:condition-optimal-S}
\end{equation}
for a unique $S^{*}\in\left\{ 1,\dots,K\right\} $. 

Using Theorem \ref{thm:solution_pk} for solving the inner problem
and the optimality condition (\ref{eq:condition-optimal-S}) for solving
the outer problem (\ref{eq:outer-subproblem}) under heavy traffic
approximation, Algorithm \ref{alg:bisection-search} summarizes the
\emph{Feedback Filtering Control Algorithm (FFCA)}, which finds the
feedback probability vector $\{p_{k}^{*}\}$ in Stage I. 

The proposed two-timescale user scheduling algorithm can be summarized
as follows. First of all, determine the optimal user feedback amount
$S^{*}$ by solving (\ref{eq:outer-subproblem}) using the FFCA. Secondly,
choose $S^{*}$ users who have the longest queues among all the $K$
users to feedback to the BS according to the policy decision $\{p_{k}^{*}(\mathbf{Q})\}$
in (\ref{eq:optimal-pk}). Thirdly, the selected users feedback their
effective SINRs based on $\{p_{k}^{*}(\mathbf{Q})\}$ and the BS schedules
the users to maximize the queue-weighted throughput as described in
the stage II policy.

Although the FFCA is derived using heavy traffic approximation, it
is in fact \emph{throughput optimal }as summarized below.
\begin{lyxThmQED}
[Throughput optimality of the FFCA]\label{thm:throughput-opt-fb-policy-1}
Suppose $\{H_{k}(t)\}$ are i.i.d. over $k$ and $t$. The feedback
control $\mathbf{p}^{*}(\mathbf{Q})$ given by FFCA achieves the maximum
stability region $\mathcal{C}$ in the MU-MIMO system. 
\end{lyxThmQED}
\begin{proof}
Please refer to Appendix \ref{app:Pf-thm-throughput-optimal} for
the proof.
\end{proof}

\begin{algorithm}
\begin{enumerate}
\item Initialization: $S:=\lfloor\frac{K}{2}\rfloor$. $S_{\min}=1$, $S_{\max}=K$.
\item \label{enu:Core-step}Evaluate the condition in (\ref{eq:condition-optimal-S}).
If $\hat{\mathcal{U}}(S^{*})\geq\hat{\mathcal{U}}(S^{*}-1)$, then
$S_{\min}:=S$. Otherwise, $S_{\max}:=S$.
\item \label{enu:Found-S*}Repeat \emph{Step \ref{enu:Core-step})} by setting
$S:=\lfloor(S_{\min}+S_{\max})/2\rfloor$, until $S_{\max}-S_{\min}\leq1$.
\item Find the optimal user feedback probability vector $\mathbf{p}$ according
to (\ref{eq:optimal-pk}) in Theorem \ref{thm:solution_pk}, by setting
$S=S^{*}$ found from \emph{Step \ref{enu:Found-S*})}. The algorithm
thus finishes.
\end{enumerate}
\caption{\label{alg:bisection-search}Feedback Filtering Control Algorithm
(FFCA)}
\end{algorithm}

\section{Large Deviation Delay Analysis for the Worst Case User\label{sec:Queueing-delay-analysis}}

In this section, we will study the queueing delay performance of the
proposed solution and illustrate the gain of having queue-aware policy.
We are interested in the steady state distribution of the worst case
queueing performance, i.e., 
\[
\lim_{t\to\infty}\mbox{Pr}(\max_{1\leq k\leq K}Q_{k}(t)>B)
\]
where $B$ is the buffer size. We denote $Q_{\max}(t)=\max_{k}Q_{k}(t)$
as the maximum queue length process and $Q_{\max}(\infty)$ as the
steady state of the $Q_{\max}(t)$. To overcome the technical challenges
associated with delay analysis of MU-MIMO system, we consider the
large deviation approach \cite{weiss1995large}. Specifically, we
focus on the asymptotic overflow probability for the maximum queue
$Q_{\max}(\infty)$ over a large buffer size $B$, which is captured
by the large deviation decay rate of the tail probability of $Q_{\max}(\infty)$.
In the next section, we shall introduce the decay rate function for
$Q_{\max}(\infty)$.

\subsection{Large Deviation Decay Rate for $Q_{\max}(\infty)$ Using Sample Path
Analysis}

The large deviation decay rate function $I^{*}$ for the tail probability
of $Q_{\max}(\infty)$ is defined as 
\begin{equation}
I^{*}\triangleq\lim_{B\to\infty}-\frac{1}{B}\log\mbox{Pr}\left(Q_{\max}(\infty)>B\right).\label{eq:I(K)-ldp-rate}
\end{equation}
Note that, with the notion of the large deviation rate function, the
queue overflow probability can be written as 
\begin{equation}
\mbox{Pr}(Q_{\max}(\infty)>B)=e^{-I^{*}B+o(B)}\label{eq:I(k)-exponential-property}
\end{equation}
where the component $I^{*}$ controls how fast the queue overflow
probability drops when the buffer size $B$ grows. A larger decay
rate $I^{*}$ corresponds to a better performance of the scheduling
algorithm in the sense of reducing the worst case delay $Q_{\max}$
in the system.  

To find the large deviation decay rate $I^{*}$, we first study the
packet departure process $D_{\max}(t)$ associated with the maximum
queue $Q_{\max}(t)$. Denote $D_{\max}(t)=R_{\max}(t,\mathbf{Q}(t))/L$,
where $R_{\max}(t,\mathbf{Q}(t))$ is the transmission data rate in
bits. Define the $\tau$-range \emph{logarithm moment generating function}
(LMF) as $\Lambda_{D}^{\tau}(\theta)=\frac{1}{\tau}\log\mathbb{E}\left[\exp\left(\theta\sum_{t=1}^{\tau}D_{\max}(t)\right)\right]$.
We consider a {}``near i.i.d.'' property for the departure process
$D_{\max}(t)$, which is captured in the following%
\footnote{A comprehensive technique to verify the assumption is given in \cite[Theorem 9.3]{Gulinsky:1993fk}.
For easy discussion, we omit the details here.%
}. 
\begin{asmQED}
[Existence of the LMF]\label{asm:existence-LMF} The limit of the
$\tau$-range LMF exists as an extended real number $\mathbb{R}\cup\{+\infty\}$
for each $\theta\in\mathbb{R}$, i.e., $\lim_{\tau\to\infty}\Lambda_{D}^{\tau}(\theta)\triangleq\Lambda_{D}(\theta)$.
\end{asmQED}

Note that, a simple example to satisfy the above assumption is $D_{\max}(t)$
being i.i.d., where $\Lambda_{D}^{\tau}(\theta)=\Lambda_{D}(\theta)=\log\mathbb{E}\left[\exp\left(\theta D_{\max}\right)\right]$. 

For easy discussion, consider i.i.d. arrivals $A_{k}(t)$ with mean
$\mathbb{E}\left[A_{k}\right]=\lambda$ and LMF $\log\psi_{A,k}(\theta)\triangleq\Lambda_{A}(\theta)$.
Denote $g(x,\theta)=\Lambda_{A}(\theta)+\Lambda_{D}(x,-\theta)$,
where $x$ represents some system state according to the scheduling
policy. We carry out a sample path analysis as follows.

Consider a scaled sample path $q_{\max}^{B}(t)=\frac{1}{B}Q_{\max}(\lfloor Bt\rfloor)$,
which starts from $q_{\max}^{B}(0)=0$ and reaches $q_{\max}^{B}(T_{s})=1$,
for some $T_{s}$. With the scaling, we have $\mbox{Pr}(Q_{\max}(\infty)>B)=\mbox{Pr}\left(q_{\max}^{B}(\infty)>1\right)$.
Let $w(t)$ be a continuous sample path following $q_{\max}^{B}(t)$,
as $w(t)\approx q_{\max}^{B}(t)$. We focus on the rate function $I_{0}$
defined as $I_{0}=$ 
\[
\inf_{w(\centerdot)}\left\{ \int_{0}^{T_{s}}l(w(\tau),w^{'}(\tau))d\tau:w(0)=0,w(T_{s})=1,T_{s}>0\right\} 
\]
where 
\begin{equation}
l(x=w(\tau),y=w^{'}(\tau))\triangleq\sup_{\theta}\left\{ \theta y-g(x,\theta)\right\} \label{eq:local-rate-func}
\end{equation}
is the \emph{local rate} function \cite{weiss1995large}. As an intuitive
illustration, $I_{0}$ corresponds to finding a {}``least cost''
path $w^{*}(t)$ that goes overflow at $w(T_{s})=1$. In other words,
the $q_{\max}^{B}(t)$ {}``most likely'' follows the path $w^{*}(t)$
to overflow, if it would.

We then connect the $I_{0}$ defined above with the large deviation
principle of $Q_{\max}(\infty)$ in the following results.
\begin{lyxThmQED}
[The large deviation principle for $Q_{\max}(\infty)$]\label{thm:LDP_rate}
Suppose $g(x,\theta)$ is Lipschitz continuous on $x\in[0,1]$. Then
\[
\lim_{B\to\infty}\frac{1}{B}\log\mathbb{E}\left[\mbox{Pr}(q_{\max}^{B}(\infty)>1)\right]=-I_{0}.
\]
In addition, assume that $l(x,y)$ in (\ref{eq:local-rate-func})
is differentiable in $y$ at all $x$, which is non-degenerate in
$[0,1]$. For each $x$, the equation $g(x,\theta^{*}(x))=0$ has
at most two solutions. Then with the appropriate choice of $\theta^{*}(x)$,
we have 
\begin{equation}
I_{0}=\int_{0}^{1}\theta^{*}(x)dx.\label{eq:LDP-rate}
\end{equation}

\end{lyxThmQED}
\begin{proof}
Please refer to Appendix \ref{app:LDP-rate-proof} for the proof.
\end{proof}

As an application example for the above result, we calculate the rate
function for a CSI-only baseline scheduling algorithm: Each user $k$
feeds back the SINR for the $i^{*}(k,n)$-th beam on each antenna
$n$, where $i^{*}(k,n)=\arg\max_{1\leq i\leq M}\mbox{SINR}_{k,n}^{i}$.
On the other hand, the BS schedules the user with the highest SINR
on each beam $i$, for $i=1,\dots,M$. Consider i.i.d Poisson arrivals
$A(t)$ with parameter $\lambda=\lambda_{tot}/K$, and i.i.d. CSI
$\{H_{k}\}$. We have the following results.

\begin{lyxCorQED}
[Decay rate for the CSI-only algorithm]\label{cor:Decay-rate-CSI-baseline}Assume
$\mu_{b}\triangleq\frac{M\log\left(P\log NK\right)}{KL}>\lambda$.
The large deviation decay rate for $Q_{\max}(\infty)$ under the CSI-only
baseline algorithm can be expressed as 
\begin{equation}
I_{\mathrm{baseline}}^{*}\approx\log\frac{M\log\left(P\log NK\right)}{\lambda_{tot}L}.\label{eq:ldp-rate-baseline}
\end{equation}
which is asymptotically accurate at large $M$ and $K$.
\end{lyxCorQED}
\begin{proof}
Please refer to Appendix \ref{app:proof-cor-decay-rate-CSI-baseline}
for the proof.
\end{proof}

The above result shows that the CSI-only baseline algorithm has a
decay rate $I_{\mathrm{baseline}}^{*}=\mathcal{O}(\log\log\log K)$.
We will show later that, by taking into account the QSI in the user
scheduling, the proposed two-timescale algorithm achieves a much larger
decay rate of the overflow probability.

\subsection{Asymptotic Data Rate of the Proposed Algorithm}

 To derive the large deviation decay rate $I^{*}$ for $Q_{\max}(t)$
under the proposed algorithm, we need to understand the corresponding
packet departure rate $D_{\max,p}(t)$. Denote $D_{\max,b}(t;S)$
as the packet departure rate under the CSI-only algorithm for a group
of $S$ users. We have the following property.
\begin{lyxLemQED}
[Property of $D_{\max,p}(t)$]\label{lem:Property-D_max} Given
$|\mathcal{F}|=S$ users feedback, we have 
\begin{equation}
D_{\max,b}(t;S)\leq D_{\max,p}(t;S)\leq\frac{1}{L}\sum_{n=1}^{N}\log(1+\mbox{SINR}_{m(t),n}^{i^{*}(n)})\label{eq:bound-D_max}
\end{equation}
where $\mbox{SINR}_{m(t),n}^{i^{*}(n)}$ is the SINR on the $n$-th
receive antenna of the $k=m(t)$ user who has the longest queue and
feeds back the $i^{*}(n)$-th beam.
\end{lyxLemQED}

The left hand side of (\ref{eq:bound-D_max}) is due to the fact that
the maximum queue user has a higher probability to get scheduled under
the Stage II queue-weighted scheduling policy. The equality holds
when all the feedback users have similar queue length, i.e., $Q_{\pi(1)}=Q_{\pi(S)}$.
The equality on the right hand side of (\ref{eq:bound-D_max}) holds
when the maximum queue user has dominating queue length, i.e., $Q_{\pi(1)}\gg Q_{\pi(2)}$,
and hence must be scheduled.

In addition, we derive the following result for evaluating the feedback
amount $S^{*}$.
\begin{lyxLemQED}
[Upper bound of $S^{*}$]\label{lem:solution-S} The upper bound
of $S^{*}(t)$ which solves (\ref{eq:outer-subproblem-2}) is given
by
\begin{equation}
S^{*}(\mathbf{Q}(t);K)\leq\min\left\{ e^{W(c_{1})}/N,K\right\} \triangleq\hat{S}^{*}(Q_{\max})\label{eq:optimal-solution-approx}
\end{equation}
where $c_{1}=\frac{MNQ_{\max}}{V}$, and $W(x)$ is the Lambert W
function \cite{corless1996lambertw} defined as $W(x)e^{W(x)}=x$.
The equality holds when $Q_{\pi(k)}\equiv Q_{\max}$ for all $k$.
\end{lyxLemQED}
\begin{proof}
Please refer to Appendix \ref{app:proof-lem-sol-S} for the proof.
\end{proof}
\begin{remrk}
[Interpretation of $S^{*}$] The results provides an important insight
that, when $Q_{\max}$ is large, it is better to have more user feedback
to boost up the system throughput. On the other hand, when $Q_{\max}$
is small, we can have less user feedback and give higher priorities
to the urgent users.  

With the results of Lemma \ref{lem:Property-D_max} and \ref{lem:solution-S},
we can obtain the packet departure rate for $Q_{\max}(t)$. We thus
study the large deviation decay rate for the proposed algorithm in
the next subsection.
\end{remrk}

\subsection{Rate Function for the Proposed Algorithm under $T=1$}

\label{sub:Asymptotic-comparison-CSI-only-baseline}

To gain more insight from the general results in Theorem \ref{thm:LDP_rate},
we consider a special case where the CSI $\{H_{k}\}$ are i.i.d.,
and the arrivals $A_{k}$ follow the Poisson distribution with parameter
$\lambda_{k}=\lambda=\lambda_{tot}/K$. 

We first consider the case $T=1$, where the BS broadcasts the updated
feedback policy $\hat{p}_{k}(\mathbf{Q})$ at every time slot. We
obtain the following results for the large deviation decay rate of
$Q_{\max}(\infty)$ under the proposed two-timescale user scheduling
algorithm.
\begin{lyxThmQED}
[Decay rate for the proposed algorithm]\label{thm:proposed-scheme-rate-function}
Let $\mu_{p}(x)=\frac{M\log\left(P\log N\hat{S}^{*}(x)\right)}{L\hat{S}^{*}(x)}$.
Assume that $\lambda<\inf_{x\in[0,1]}\mu_{p}(x)$. Then the large
deviation decay rate of $Q_{\max}(\infty)$ under the two-timescale
user scheduling algorithm can be expressed as 
\begin{equation}
I_{\mathrm{prop}}^{*}\geq\left(1-\epsilon\right)\log K+\log\frac{M}{\lambda_{tot}L}+\epsilon\log r_{0}+C\triangleq I_{\mathrm{prop}}^{LB}\label{eq:ldp-rate-proposed}
\end{equation}
where $\epsilon>0$ is a small constant, $r_{0}=\int_{0}^{1}\log\left(1+x\right)dF(x)$,
and $C=\int_{\epsilon}^{1}\left\{ \log\left[N\log\left(PW\left(\frac{MNx}{V}\right)\right)\right]-W\left(\frac{MNx}{V}\right)\right\} dx$.
\end{lyxThmQED}
\begin{proof}
Please refer to Appendix \ref{app:LDP-rate-proposed-scheme} for the
proof.
\end{proof}

Based on the results in Corollary \ref{cor:Decay-rate-CSI-baseline}
and Theorem \ref{thm:proposed-scheme-rate-function} we conclude the
following for the CSI-only user scheduling algorithm and the proposed
two-timescale algorithm.
\begin{itemize}
\item \emph{Gain of the queue-aware policy:} Large deviation decay rates
$I_{\mathrm{prop}}^{*}\gg I_{\mathrm{baseline}}^{*}$, when the number
of users $K$ grows large. This demonstrates that it is important
to utilize the queue information in the user scheduling algorithm
to minimize  the worst case delay.
\item \emph{Impact of the multi-user diversity:} In addition, both of the
schemes benefit from the increase of the number of users $K$, as
seen from the terms $\log(P\log NK)$ in (\ref{eq:ldp-rate-baseline})
and $\log(K)$ in (\ref{eq:ldp-rate-proposed}). The decay rate increases
when the number of users increases, and the rate $I_{\mathrm{prop}}^{*}$
increases faster than the baseline.
\item \emph{Impact of the multi-antenna transmission:} Furthermore, both
of the schemes benefit from the MU-MIMO channel. It is demonstrated
that, when increasing the number of data streams $M$ and the receive
antennas $N$, the large deviation decay rates $I_{\mathrm{prop}}^{*}$
and $I_{\mathrm{baseline}}^{*}$ both increase as $\mathcal{O}(\log M\log\log N)$.
\end{itemize}

In summary, by carefully exploiting the queue information in the stage
I feedback filtering, the proposed MU-MIMO algorithm has significant
delay performance gain compared with conventional CSI-only schemes.

\subsection{Rate Function for $T>1$}

Now we consider the $T$-step feedback policy, where the BS updates
the $\hat{p}_{k}(\mathbf{Q})$ for every $T>1$ time slot. Denote
the corresponding maximum queue process as $Q_{\max}^{(T)}(t)$. We
are interested in the case where the process $Q_{\max}^{(T)}(t)$
is stable and assume the large deviation principle exists. 

Define the rate function as 
\[
I_{\mathrm{prop}}^{(T)*}\triangleq\lim_{B\to\infty}-\frac{1}{B}\log\mbox{Pr}\left(Q_{\max}^{(T)}(\infty)>B\right).
\]
For easy discussion, we consider i.i.d. arrivals $A_{k}(t)$ and i.i.d.
CSI $\{H_{k}(t)\}$. Consider a random process $v(t)=A_{1}(t)-A_{2}(t)-d(t)$,
where $A_{1}$ and $A_{2}$ are two i.i.d. arrival sequences, $d(t)$
has probability distribution function given by $F(P^{-1}(2^{x}-1))$
and $F(x)$ is defined in (\ref{eq:snr_cdf}). We have the following
result for the decay rate of the $T$-step feedback policy.
\begin{lyxThmQED}
[Decay rate for the $T$-step feedback policy]\label{thm:rate-function-T-step}
Assume the conditions in Theorem \ref{thm:proposed-scheme-rate-function},
we have 
\[
I_{\mathrm{prop}}^{(T)*}\geq I_{\mathrm{prop}}^{LB}-\int_{0}^{1}\rho(x)dx
\]
where $\rho(x)\triangleq-\frac{1}{\hat{\mu}_{p}(x)-\lambda}\log\left(e^{\hat{\mu}_{p}(x)-\lambda}-(e^{\hat{\mu}_{p}(x)-\lambda}-1)P_{0}^{T}\right)$
and $P_{0}^{T}\triangleq\mbox{Pr}\big\{\sum_{\tau=1}^{T-1}v(\tau)>0\big\}$.
\end{lyxThmQED}
\begin{proof}
Please refer to Appendix \ref{app:proof-rate-function-T-step} for
the proof.\end{proof}
\begin{remrk}
[Impact of $T$ and the arrival distribution] Note that $P_{0}^{T}$
represents a lower bound probability for the maximum queue user remaining
in the outdated feedback group $\mathcal{F}(t_{0})$ during $t\in[t_{0},t_{0}+T)$;
the larger the $T$, the smaller the $P_{0}^{T}$. The lower bound
becomes tight when $P_{0}^{T}$ is close to $1$. The above result
shows that the decay rate function $I_{\mathrm{prop}}^{(T)*}$ decreases
when the QSI update period $T$ increases. Moreover, the distribution
of arrival plays an important role in $T>1$. With a heavier tail
for the arrival, $P_{0}^{T}$ decreases, resulting in a higher performance
penalty for $T>1$. Finally, the performance in terms of the overflow
probability for the two-timescale algorithm is sensitive to the timely
queue-aware feedback under heavy loading when $\hat{\mu}_{p}-\lambda$
is small.~\hfill\IEEEQED
\end{remrk}

\section{Numerical Results\label{sec:Numerical-Results}}

In this section, we simulate the queueing delay performance of the
proposed two-timescale user scheduling algorithm. We consider a MU-MIMO
system with $K$ users, and packets arrive to the queue of each user
according to a Poisson distribution with rate $\lambda=\lambda_{tot}/K$,
where the total arrival rate is $\lambda_{tot}=7500$ packets/second.
Each packet has $L=8000$ bits. The system bandwidth is $10$ MHz
and the SNR is 10 dB. The number of transmit and receive antennas
are $M=4$ and $N=2$, respectively. The scheduling time slot is $\tau=1$
ms and the simulation is run over $T_{tot}=100$ seconds. We compare
the performance of proposed algorithm against the following reference
baselines.
\begin{itemize}
\item \textbf{Baseline 1: CSI-only user scheduling (CSIO)} \cite{zhang2007mimo}.
At each time slot, all the users feedback the CSI to the BS, and the
BS schedules a set of users who respectively have the highest SINR
on each beam (see Section \ref{sub:Asymptotic-comparison-CSI-only-baseline}). 
\item \textbf{Baseline 2: CSI-only user scheduling with limited feedback
(CSIO-LF)} \cite{zhang2007mimo}. The scheme is similar to baseline
1 except that the user feeds back to the BS only when its SINR exceeds
a threshold $t_{SINR}=1$ dB.
\item \textbf{Baseline 3: Proportional fair user scheduling (PFS)} \cite{yoo2006optimality}.
At each time slot, all the users feedback the CSI to the BS, and the
BS transmits data to the users using proportional fair scheduling
with window size $t_{w}=100$ ms.
\item \textbf{Baseline 4: Max weighted queue user scheduling (MWQ)} \cite{neely2005dynamic}.
At each time slot, all the users feedback their CSI to the BS, and
the BS selects a set of users so that the instantaneous queue-weighted
sum rate $\sum Q_{k}R_{k}$ is maximized.
\end{itemize}
Note that the associated user scheduling problem in baseline 4 has
much higher complexity for user scheduling and feedback from all the
users are required. Hence, baseline 4 serves for performance benchmarking
purpose only. %

\subsection{Queueing Performance and Feedback Comparisons}

\begin{figure}
\begin{centering}
\includegraphics[width=1\columnwidth]{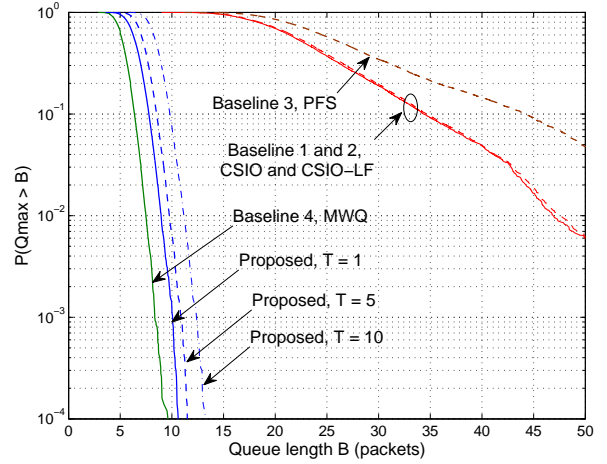}
\par\end{centering}

\caption{\label{fig:overflow-prob}The overflow probability for the worst case
queue $\mbox{Pr}\left(Q_{\max}(\infty)>B\right)$ versus the buffer
size $B$. The number of users is $K=40$. The feedback policy $\bm{\chi}$
in stage I updates on every $T=1,5,10$ time slots. The proposed scheme
significantly outperforms over baselines 1 - 3. It also performs closely
to baseline 4.}
\end{figure}

\begin{figure}
\begin{centering}
\includegraphics[width=1\columnwidth]{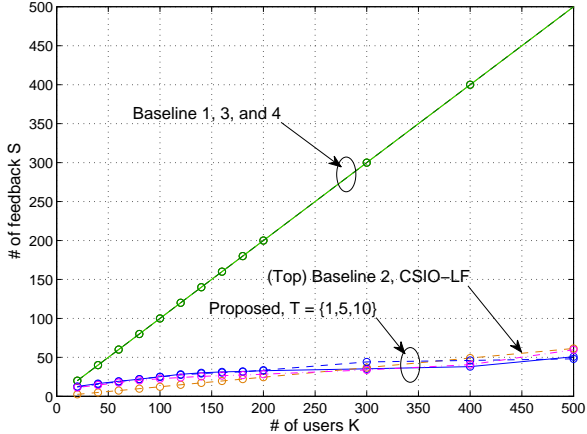}
\par\end{centering}

\caption{\label{fig:feedback-comp}The average feedback amount $\overline{S}$
versus the number of users $K$. The feedback threshold of baseline
2 is $t_{SINR}=1$ dB. The feedback amount of the proposed scheme
is much less than those of all the baselines. Note that although baseline
4 (MWQ) has a smaller worst case queue, it requires all the users
feedback to the BS.}
\end{figure}

Fig. \ref{fig:overflow-prob} shows the overflow probability for the
worst case queue $\mbox{Pr}\left(Q_{\max}(\infty)>B\right)$ versus
the buffer size $B$. The number of users is $K=40$. The feedback
policy $\bm{\chi}$ updates on every $T=1,5,10$ time slots. The proposed
scheme significantly outperforms over baselines 1 - 3. It also has
similar performance as baseline 4. Fig. \ref{fig:feedback-comp} demonstrates
the average feedback amount $\overline{S}$ (defined as the average
number of users feedback to the BS at each time slot) versus the number
of users $K$. The feedback amount of the proposed scheme is less
than those of all the baselines. Note that although baseline 4 has
a smaller worst case queue, it requires all the users feedback to
the BS.

\subsection{Large Deviation Decay Rate for a Large Number of Users}

\begin{figure}
\begin{centering}
\includegraphics[width=1\columnwidth]{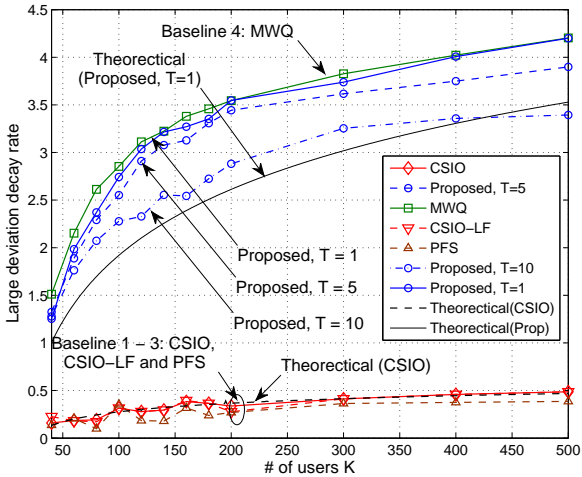}
\par\end{centering}

\caption{\label{fig:decay-rate-over-K}The large deviation decay rate over
the number of users. The decay rate for the proposed scheme grows
much faster than that of baselines 1 - 3 with the number of users
$K$. Note that although baseline 4 performs the best, it requires
all the users feedback to the BS.}
\end{figure}

Fig. \ref{fig:decay-rate-over-K} shows the large deviation decay
rate over the number of users. The decay rate for the proposed scheme
grows much faster than those of baselines 1 - 3 with the number of
users $K$. Moreover, the theorectical rate functions are plotted.
These are consistent with the results in Corollary \ref{cor:Decay-rate-CSI-baseline}
and Theorem \ref{thm:proposed-scheme-rate-function}.

\section{Conclusions\label{sec:Conclusions}}

In this paper, we proposed a novel two-timescale delay-aware user
scheduling algorithm for the MU-MIMO system. The policy consists of
a queue-aware mobile-driven feedback filtering stage and a dynamic
queue-weighted user scheduling stage. The queue-aware feedback filtering
control algorithm in stage I was derived through solving an optimization
problem. Under the proposed two-timescale user scheduling algorithm,
we also evaluated the queueing delay performance for the worst case
user using the sample path large deviation analysis. The large deviation
decay rate for the proposed algorithm, scaled as $\mathcal{O}\left(\log K\right)$,
was shown to be much larger than a CSI-only user scheduling algorithm,
which means that the proposed scheme performs better in reducing the
worst case delay. The numerical results demonstrated a significant
performances gain over the CSI-only algorithm and a huge feedback
reduction over the MWQ algorithm. 

\appendices

\section{Poof of Theorem \ref{thm:solution_pk} \label{app:Poof-thm-solution_pk}}

Note that the amount of feedback $s=\sum_{k}\chi_{k}$ follows the
Poisson Binomial distribution, which is insensitive of individual
$p_{k}$ given a fixed $\sum_{k}p_{k}=S$ \cite{Hong:2011uq}. For
an easy elaboration, consider a Poisson distribution (which is close
to the Poisson Binomial distribution) with parameter $\sum_{k}p_{k}=S$
to approximate the distribution of $s$. The approximation error is
upper bounded by $2\sum_{k}p_{k}^{2}$ \cite{Hong:2011uq}. 

We first find the optimal solution under the heavy traffic approximation,
and then we generalize the result into the normal case. In the heavy
traffic case where $Q_{\pi(1)}\approx Q_{\pi(K)}$, the objective
in (\ref{eq:inner-subproblem}) can be written as $f(\mathbf{p})=\sum_{k}Q_{k}(t)\mathbb{E}\left[\chi_{k}\eta(s)\right]=\sum_{k}Q_{k}\mathbb{E}\big\{\mathbb{E}[\chi_{k}\eta(s)\big|\chi_{k}]\big\}\approx\sum p_{k}Q_{k}\mathbb{E}\eta(s)$,
where $\mathbb{E}[\chi_{k}\eta(s)\big|\chi_{k}]=p_{k}\mathbb{E}\eta(s)+o(\sum_{k}p_{k})\approx p_{k}\mathbb{E}\eta(s)$,
and $\eta(s)$ does not depend on $\mathbf{Q}$ since all $Q_{k}$
are almost the same. Thus $\mathbb{E}\eta(s)$ can be computed by
an approximated Poisson distribution which does not depend on $\chi_{k}$.

As such, the inner subproblem becomes a linear program with constraints
$\sum p_{k}\leq S$ and $0\leq p_{k}\leq1$, $\forall k$. The solution
is given by $p_{\pi(k)}=1,\;1\leq k\leq\lfloor S\rfloor$, $p_{\pi(k_{0})}=S-\lfloor S\rfloor,\; k_{0}=\lfloor S\rfloor+1$,
and $p_{\pi(k)}=0$, otherwise, where the permutation $\Pi=\{\pi(k)\}$
is such that $Q_{\pi(1)}\geq\dots\geq Q_{\pi(K)}$. 

Now we show that the above solution is also a local optimum under
general queueing profiles. Consider an arbitrary feasible probability
vector $\mathbf{\widetilde{p}}=\mathbf{p}^{*}+\mathbf{p}^{\epsilon}$
lies in a small neighborhood of $\mathbf{p}^{*}$. Since $\sum_{k}\widetilde{p}_{k}=S$,
we must decrease a probability of $p_{0}^{\epsilon}$ for some user
$k=\pi(j)$, $j\leq S$, in order to increase a probability $p_{0}^{\epsilon}$
for a user $k^{'}=\pi(j^{'})$, $j^{'}>S$. The differential utility
$\mathcal{W}(\widetilde{\mathbf{p}};S)-\mathcal{W}(\mathbf{p};S)$
then becomes 
\begin{eqnarray*}
\triangle\mathcal{W}(S) & = & -p_{0}^{\epsilon}Q_{k}\mathbb{E}[R_{k}\big|Q_{k}R_{k}\in\max^{M}\{Q_{i}R_{i},i\in\mathcal{F}\}]\\
 &  & \quad\times\mbox{Pr}(Q_{k}R_{k}\in\max^{M}\{Q_{i}R_{i},i\in\mathcal{F}\})\\
 &  & \qquad+p_{0}^{\epsilon}Q_{k^{'}}\mathbb{E}[R_{k^{'}}\big|Q_{k^{'}}R_{k^{'}}\in\max^{M}\{Q_{i}R_{i},i\in\mathcal{F}\}]\\
 &  & \qquad\times\mbox{Pr}(Q_{k^{'}}R_{k^{'}}\in\max^{M}\{Q_{i}R_{i},i\in\mathcal{F}\})
\end{eqnarray*}
where $\max^{M}\{A\}$ means a subset of $A$ with $M$ elements which
are the largest. Since $Q_{k}\geq Q_{k}^{'}$, and $R_{k}$ and $R_{k^{'}}$
are identical, we must have $\mbox{Pr}(Q_{k}R_{k}\in\max^{M}\{Q_{i}R_{i},i\in\mathcal{F}\})]\geq\mbox{Pr}(Q_{k^{'}}R_{k^{'}}\in\max^{M}\{Q_{i}R_{i},i\in\mathcal{F}\})]$.
Therefore, the differential utility cannot be positive. As $\mathbf{p}^{\epsilon}$
can be arbitrary, the vector $\mathbf{p}^{*}$ must achieve the local
maximum utility.

Moreover, as the inner problem is a GP, $\mathbf{p}^{*}$ is also
a global optimum.

\section{Proof of Lemma \ref{lem:data-rate-strongest-user}}

\label{app:maximum-data-rate}

Consider $Q_{\pi(1)}\approx Q_{\pi(S)}$. The queue weighted user
scheduling algorithm degenerates to a max-SINR based algorithm. Then
the order statistics can be applied to study the expected data rate,
and each user has around $1/S$ probability to be scheduled independently
on each beam.

From the effective SINR expression in (\ref{eq:effective-SINR}),
as $\phi_{i}$ are unitary vectors, $|H_{k}^{(n)}\phi_{i}|^{2}$ are
i.i.d. over $i$ with chi-square distribution with degrees of freedom
$2$. Consequently, the term $\sum_{j:j\neq i}\left|H_{k}^{(n)}\phi_{j}\right|^{2}$
is chi-square distributed with degrees of freedom $2M-2$. Thus, the
PDF $f(x)$ and CDF $F(x)$ of $\mbox{SINR}_{k,n}^{i}$ are given
by $f(x)=\frac{e^{-x/P}}{(1+x)^{M}}\left(\frac{1}{P}(1+x)+M-1\right)$
and $F(x)=1-\frac{e^{-x/P}}{(1+x)^{M-1}}$, respectively \cite{sharif2005capacity}.
Thus, for a particular user $k\in\mathcal{F}$, as $\mbox{SINR}_{k,n}^{i}$
are i.i.d. over different users $k$ and antennas $n$, the probability
that user $k$ has the largest SINR on the $i$-th beam and the $n$-th
antenna is give by $1/NS$. The corresponding CDF of the maximum SINR
is 
\begin{eqnarray}
P\left(\max_{k\in\mathcal{F},1\leq n\leq N}\mbox{SINR}_{k,n}^{i}\leq x\right) & = & \left(F(x)\right)^{NS}\label{eq:max_sinr-cdf}
\end{eqnarray}
and hence, the data rate can be given by 
\begin{eqnarray*}
\hat{R} & = & \int_{0}^{\infty}\log(1+x)d(F(x))^{NS}\\
 & = & \int_{0}^{\infty}\log(1+x)NSf(x)F(x)^{NS-1}dx.
\end{eqnarray*}

As each user equips with $N$ antennas, the average data rate for
user $k\in\mathcal{F}$, given $|\mathcal{F}|=S$ is $\eta_{k}(S)\approx\sum_{n=1}^{N}\sum_{i=1}^{M}\mbox{Pr}\left(\mbox{SINR}_{k,n}^{i}=\max_{k_{0}\in\mathcal{F},1\leq n\leq N}\mbox{SINR}_{k_{0},n}^{i}\right)\hat{R}=NM\frac{1}{NS}\hat{R}=\hat{\eta}(S)$.

\section{Proof of Theorem \ref{thm:outer-concavity-integer}}

\label{app:proof-thm-concave-integer}

We first note that the function $\hat{\mathcal{W}}(S)$ is piece-wise
linear and so does $\hat{\mathcal{U}}(S)$. Then the function $\hat{\mathcal{U}}(S)$
is concave if we can find a a smooth and concave upper envelope function
that passes through every corner point of $\hat{\mathcal{U}}(S)$.

Let $\mathcal{I}$ denote the space of twice-differentiable positively
non-decreasing concave functions, i.e., $\mathcal{I}\triangleq\left\{ \phi\in\mathcal{C}^{2}(0,+\infty):\phi>0,\phi^{'}\geq0,\phi^{''}\leq0\right\} .$
Let $\eta_{c}(s)=\hat{\eta}(s)$, where $\eta_{c}(s)$ is allowed
to take real values. Given $g\in\mathcal{I}$, define $G(s)=g(s)\eta_{c}(s)-Vs$.
We have the following result.
\begin{lyxLemQED}
\label{lem:sufficient-condition-local-maximum} $G(s)$ is concave
for any $g\in\mathcal{I}$.
\end{lyxLemQED}
\begin{proof}
To show $G(s)$ is concave is equivalent to showing $G^{''}(s)=g^{''}(s)\eta_{c}(s)+2g^{'}(s)\eta_{c}^{'}(s)+g(s)\eta_{c}^{''}(s)\leq0$. 

From the property of $g\in\mathcal{I}$, we have $g^{'}(s)s\leq g(s)$.
Thus 
\begin{equation}
G^{''}(s)\leq g^{''}(s)\eta_{c}(s)+\frac{g(s)}{s}\left[2\eta_{c}^{'}(s)+s\eta_{c}^{''}(s)\right].\label{eq:dd_G(s)}
\end{equation}
The first term is negative by the definition of $g\in\mathcal{I}$.
In the second term, $\frac{g(s)}{s}$ is positive. Now, let $\Gamma(s)=2\eta_{c}^{'}(s)+s\eta_{c}^{''}(s)$.
Note that, from (\ref{eq:data-rate-max-user}), $\eta_{c}(s)$ is
twice differentiable on $s\in(0,+\infty)$, and we have the following
two equations 
\[
\eta_{c}^{'}(s)=M\int_{0}^{\infty}\log(1+x)N^{2}f(x)\log[F(x)]F(x)^{NS-1}dx,
\]
\[
\eta_{c}^{''}(s)=M\int_{0}^{\infty}\log(1+x)N^{3}f(x)\log\left[F(x)\right]^{2}F(x)^{NS-1}dx.
\]

One can easily verify that, $\Gamma(s;N=1)\leq0$ for all $s>0$.
This can be seen by first numerically verifying $\Gamma(s;N=1)<0$
for small $s$ (e.g., $s<1000$), and then verifying $\Gamma(s)^{'}>0$
for large $s$ through analyzing the dominating components $F(x)^{S-1}$
in the integrand as $F(x)$ sufficiently close to 1. Moreover, for
$s\to\infty$, $\Gamma(s;N=1)\to0$. 

For $N>1$, let $t=Ns$. From the above two equations, we have $\Gamma(s;N)=N^{2}\Gamma(t;N=1)\leq0$.
 With $\Gamma(s)\leq0$, we have $G^{''}(s)\leq0$ in (\ref{eq:dd_G(s)}).
Hence $G(s)$ is concave. 
\end{proof}

Now notice that the sequence $\sum_{k=1}^{S}Q_{\pi(k)}$ is non-decreasing
for $S=1,\dots,K$, and the increment is non-increasing. Then there
must exist a function $g_{Q}\in\mathcal{I}$, such that $g_{Q}(s)$
passes throughput every point of the sequence $\sum_{k=1}^{S}Q_{\pi(k)}$,
i.e., $g_{Q}(S)=\sum_{k=1}^{S}Q_{\pi(k)}$ for $S=1,\dots,K$. According
to Lemma \ref{lem:sufficient-condition-local-maximum}, the function
$G_{Q}(s)\triangleq g_{Q}(s)\eta_{c}(s)-Vs$ is concave. Moreover,
$G_{Q}(s)$ is an upper envelope function that passes throughput every
corner point of $\hat{\mathcal{U}}(S)$. This proves that $\hat{\mathcal{U}}(S)$
is concave. 

To show the optimal solution appears at one the integer point, we
take derivative of $\hat{\mathcal{U}}(S)$ and obtain 
\[
\frac{d}{dS}\hat{\mathcal{U}}(S)=-\sum_{k=1}^{\lfloor S\rfloor}Q_{\pi(k)}\hat{\eta}(\lfloor S\rfloor)\quad+\sum_{k=1}^{\lfloor S\rfloor+1}Q_{\pi(k)}\hat{\eta}(\lfloor S\rfloor+1)-V.
\]
 It is observed that, given any integer $S_{0}$, the gradient $\frac{d}{dS}\hat{\mathcal{U}}(S)$
remains constant for any $S\in(S_{0},S_{0}+1)$. If $\frac{d}{dS}\hat{\mathcal{U}}(S)=0$,
we can consider $S_{0}$ or $S_{0}+1$ to be the local maximum. If
$\frac{d}{dS}\hat{\mathcal{U}}(S)\neq0$, using the optimality condition
\cite{Boyd:2004kx}, $S\in(S_{0},S_{0}+1)$ cannot be the maximum.
It concludes that, the maximum should be an integer.

\section{Proof of Theorem \ref{thm:throughput-opt-fb-policy-1} }

\label{app:Pf-thm-throughput-optimal}

Consider the queue dynamic in (\ref{eq:queue-dynamic}). By squaring
the equation on both sides and using the property $\left[\max\{0,x\}\right]^{2}\leq x^{2}$,
we obtain $\forall k$, 
\begin{equation}
Q_{k}^{2}(t+1)\leq Q_{k}^{2}(t)+\mu_{k}^{2}(t)-2Q_{k}(t)(D_{k}(t)-A_{k}(t))+A_{k}^{2}(t)\label{eq:queue-dynamic-inequality}
\end{equation}
Following the definition of conditional Lyapunov drift $\triangle L(\mathbf{Q}(t))$
in (\ref{eq:lyapunov-drift}), taking conditional expectations and
summing over all $k$ inequalities in (\ref{eq:queue-dynamic-inequality})
yields 
\begin{eqnarray}
\triangle L(\mathbf{Q}(t)) & \leq & \mathbb{E}\left[\sum_{k}\mu_{k}^{2}(t)+A_{k}^{2}(t)|\mathbf{Q}(t)\right]\label{eq:Lyapunov-drift-1}\\
 &  & \quad-2\sum_{k}Q_{k}(t)\mathbb{E}\left[D_{k}(t)-A_{k}(t)|\mathbf{Q}(t)\right].\nonumber 
\end{eqnarray}

Denote positive constants $\overline{\mu}_{\max}^{2}$ and $\overline{\lambda}_{\max}^{2}$
such that $\mathbb{E}\left[D_{k}^{2}(t)|\mathbf{Q}(t)\right]\leq\overline{\mu}_{\max}^{2}$
and $\mathbb{E}\left[A_{k}^{2}(t)|\mathbf{Q}(t)\right]\leq\overline{\lambda}_{\max}^{2}$.
Let $C_{0}=\overline{\mu}_{\max}^{2}+\overline{\lambda}_{\max}^{2}$.
Adding $V\mathbb{E}\left\{ \mathcal{S}(\mathbf{Q}(t)|\mathbf{Q}(t)\right\} $
on both sides, the drift (\ref{eq:Lyapunov-drift-1}) is bounded by

\begin{eqnarray}
\triangle L(\mathbf{Q}(t))+V\mathbb{E}\left\{ \mathcal{S}(\mathbf{Q}(t)|\mathbf{Q}(t)\right\} \leq C_{0}K+2\sum_{k}Q_{k}(t)\lambda_{k}\label{eq:Lyapunov-drift-2-1}\\
-2\sum_{k}Q_{k}(t)\mathbb{E}\left[D_{k}(t)|\mathbf{Q}(t)\right]+V\overline{\mathcal{S}}.\nonumber 
\end{eqnarray}

Suppose now that the arrival $\bm{\lambda}=(\lambda_{1},\dots,\lambda_{K})$
is strictly interior to the stability region $\mathcal{C}$ such that
$\bm{\lambda}+\epsilon\mathbf{1}\in\mathcal{C}$, for $\epsilon>0$.
Since channel states are i.i.d. over time slots, using the result
in \cite[Corollary 1]{neely2006energy}, it follows that there exists
a stationary randomized feedback control policy that schedules user
to feedback independent of queue $\mathbf{Q}(t)$ and yields $\mathbb{E}\left[D_{k}(t)|\mathbf{Q}(t)\right]=\mathbb{E}\left[R_{k}(t)\right]\geq\lambda_{k}+\epsilon$
and $\mathbb{E}\left[\mathcal{S}(\mathbf{Q}(t)|\mathbf{Q}(t)\right]=\overline{\mathcal{S}}(\epsilon)$.
Because the stationary policy is simply a particular feedback policy
and note that the FFCA maximizes the term $\sum_{k}\mathbb{E}\left[Q_{k}(t)R_{k}(t)\right]$
under and approximated feedback cost $\hat{S}\leq K$, the right hand
side of (\ref{eq:Lyapunov-drift-2-1}) under FFCA is thus upper bounded
by $C_{0}K-2\epsilon\sum_{k}Q_{k}(t)+VK$. 

Using the results in Lemma \ref{lem:Lyapunov-drift-stability}, it
follows that $\sum_{k}Q_{k}(t)\leq\frac{C_{0}K+V\overline{\hat{\mathcal{S}}}}{2\epsilon}\leq\frac{C_{0}K+VK}{2\epsilon}<\infty$,
which proves that the FFCA policy stabilizes all the queues.

\section{Proof of Theorem \ref{thm:LDP_rate}}

\label{app:LDP-rate-proof}

Consider the scaled sample path $q_{\max}^{B}(t)=\frac{1}{B}Q_{\max}\left(\lfloor Bt\rfloor\right)$,
where the jumps can be given by%
\footnote{Here, for easy discussion, we assume the identity $q_{\max}^{B}(\tau+\frac{1}{B})-q_{\max}^{B}(\tau)=\frac{1}{B}A_{m(\tau)}-\frac{1}{B}D_{m(\tau)}$
holds on the boundary, where the maximum queue index changes, i.e.,
$m(\tau)\neq m(\tau+\frac{1}{B})$. Note that, with the fluid approximation,
such boundary effect (which violates the above equality) vanishes
in the scaled sample path $q_{\max}^{B}$ when $B$ becomes large
(and hence the jumps becomes smaller).%
} $q_{\max}^{B}(t)-q_{\max}^{B}(t_{0})$ 
\begin{eqnarray*}
 & = & \frac{1}{B}\sum_{s=\left\lfloor Bt_{0}\right\rfloor }^{\left\lfloor Bt\right\rfloor }A_{m(s)}(s)-\frac{1}{B}\sum_{s=\left\lfloor Bt_{0}\right\rfloor }^{\left\lfloor Bt\right\rfloor }D_{m(s)}(s)
\end{eqnarray*}
for $0\leq t_{0}<t\leq T_{s}$, where $m(s)=\arg\max Q_{k}(s)$ denotes
the index of the maximum queue at time $s$. Note that, for $|t-t_{0}|$
small, the jump $q_{\max}^{B}(t)-q_{\max}^{B}(t_{0})$ is a sum of
sequence of random variables $v(s)=A_{m(s)}-D_{m(s)}$, whose $\tau$-step
LMF is given by 
\begin{eqnarray*}
\Lambda_{v}^{\tau} & = & \frac{1}{\tau}\log\mathbb{E}\left[\exp\left(\theta\sum_{s=t}^{t+\tau}\left(A_{m(s)}-D_{m(s)}\right)\right)\right]\\
 & = & \log\mathbb{E}[\exp(\theta A)]+\frac{1}{\tau}\log\mathbb{E}\left[\exp\left(-\theta\sum_{s=t}^{t+\tau}D_{m(s)}\right)\right]
\end{eqnarray*}
Under Assumption \ref{asm:existence-LMF}, taking $\tau\to\infty$,
we obtain $\Lambda_{v}^{\tau}\to g(x,\theta)$, which defines the
local rate function in (\ref{eq:local-rate-func}).

Thus one can use the Gartner-Ellis theory \cite[Theorem 2.3.6]{dembo2009large}
to show the large deviation principle associated with the local rate
function (\ref{eq:local-rate-func}) for the non-i.i.d. sequence $v(t)$
on each $(w(t),w^{'}(t))$ pair following the path $w(t)$. Then we
consider the escape time $\tau_{B}=\inf\{t>0:q_{\max}^{B}(t)>1\}$.
Using the Freidlin-Wentzell theory \cite[Theorem 6.17]{shwartz1995large},
we thus obtain the large deviation principle $\lim_{B\to\infty}\frac{1}{B}\log\mathbb{E}\left[\tau_{B}\right]=I_{0}$
for the random process $q_{\max}^{B}(t)$.

Note that the mean escape time $\tau_{B}$ implies the steady state
probability for $q_{\max}^{B}(\infty)$ staying in the set $\{q_{\max}^{B}(\infty)>1\}$,
i.e., $\lim_{B\to\infty}\frac{1}{B}\log\mathbb{E}\left[\tau_{B}\right]=\lim_{B\to\infty}-\frac{1}{B}\log\mbox{Pr}\left(q_{\max}^{B}(\infty)>1\right)$.
Therefore, the first part of the theorem is established.

The second part of the theorem completely follows \cite[Lemma C.9]{weiss1995large}
and thus we omit the details here.

\section{Proof of Corollary \ref{cor:Decay-rate-CSI-baseline}}

\label{app:proof-cor-decay-rate-CSI-baseline}

For the $i$-th beam, the CSI-only algorithm selects the user with
the highest SINR for transmission. Denote $R_{b}^{(i)}$ as the corresponding
transmission data rate. We have $\mathbb{E}R_{b}^{(i)}=K\hat{\eta}(K)$,
where $\hat{\eta}(\centerdot)$ is given in (\ref{eq:data-rate-max-user}). 

Note that we have $D_{k}=\sum_{i}^{\iota}R_{b}^{(i)}/L$, where $\iota=0,\dots,\min\left\{ M,N\right\} $
is the number of beams assigned to user $k$ and $\mathbb{E}D_{k}=\frac{M\hat{\eta}(K)}{L}\triangleq\mu_{b}$.
Since $\mbox{SINR}_{k,n}^{i}$ are i.i.d. over $k$ and $n=1,\dots,N$,
the probability for a user being assigned $\iota$ beams approximately
follows a binomial distribution $\mathcal{B}(M,p)$, with $p=\frac{1}{K}$.
  It is well-known that $\mathcal{B}(M,p)\to\mbox{Poiss}\left(\rho\right)$
with $\rho=\frac{M}{K}$, as $M,K\to\infty$. Therefore, $D_{k}$
approximately follows the distribution of 
\begin{equation}
\hat{D}_{k}(K)=\frac{\xi}{L}K\hat{\eta}(K)\label{eq:departure-rate-baseline}
\end{equation}
where $\xi\sim\mbox{Poiss}\left(\rho\right)$. The LMF of $\hat{D}_{k}$
can be easily obtained as $\Lambda_{\hat{D}}(\theta)=\mu_{b}(e^{\theta}-1)$.
Note that $Q_{\max}(t)$ and $Q_{k}(t)$ are identical under the CSI-only
algorithm. Therefore, we have an explicit expression of the LMF as
\[
g(x,\theta)=\Lambda_{A}(\theta)+\Lambda_{D}(x,-\theta)=\lambda(e^{\theta}-1)+\mu_{b}\left(e^{-\theta}-1\right).
\]
Using Theorem \ref{thm:LDP_rate} and solving $g(x,\theta)=0$, we
obtain $e^{\theta}=1$ and $e^{\theta}=\frac{\mu_{b}}{\lambda}$.
One can verify that $e^{\theta}=1$ yields trivial solution $I^{*}=0$.
Then we have 
\begin{equation}
I_{\mathrm{baseline}}^{*}\approx\log\frac{\mu_{b}}{\lambda}=\log\frac{MK\hat{\eta}(K)}{\lambda_{tot}L}.\label{eq:I-baseline-exact}
\end{equation}
Moreover, using the extreme value theorem, we obtain $\mathbb{E}R_{b}^{(i)}/\log\left(P\log NK\right)\to1$,
as $K\to\infty$ \cite{sharif2005capacity}, which implies $K\hat{\eta}(K)\to\log\left(P\log NK\right)$.
Therefore, we further have $I_{\mathrm{baseline}}^{*}\approx\log\frac{M\log\left(P\log NK\right)}{\lambda_{tot}L}.$
The conditions of Theorem \ref{thm:LDP_rate} are satisfied when $\mu_{b}>\lambda$,
or approximately, $\hat{\mu}_{b}\triangleq\frac{M\log\left(P\log NK\right)}{KL}>\lambda$.

\section{Proof of Lemma \ref{lem:solution-S} \label{app:proof-lem-sol-S}}

Consider an upper bound ordered queue length profile as follows, $\hat{Q}_{\pi(1)}=Q_{\max}$
and $\quad\hat{Q}_{\pi(j)}=Q_{\max}(1-\delta\frac{j-1}{K})$, where
$\delta\geq0$ is chosen such that $Q_{\pi(j)}\leq\hat{Q}_{\pi(j)}$
for all $j=\{1,\dots,K\}$. 

We first note that using the extreme value theorem, we have $K\hat{\eta}(K)/\log\left(P\log NK\right)\to1$,
as $K\to\infty$ \cite{sharif2005capacity}, which implies that $\hat{\eta}(K)\to\frac{M}{K}\log\left(P\log NK\right)$.
Focusing on large $K$, we may typically obtain a large $S^{*}$ which
can validate the asymptotic approximation of $\hat{\eta}(S)$. Thus
we solve the outer subproblem (\ref{eq:outer-subproblem-2}) by substituting
$Q_{\pi(k)}$ with $\hat{Q}_{\pi(k)}$ and $\eta_{\pi(k)}(S)\approx\frac{M}{S}\log\left(P\log NS\right)$
as follows,
\[
\max_{\hat{S}}\; g(\hat{S})=\frac{Q_{\max}}{2K}(2K+\delta-\delta\hat{S})M\log\left(P\log N\hat{S}\right)-V\hat{S}.
\]

It can be shown that $g(\hat{S})$ is concave. Taking derivative of
$g(\hat{S})$, and setting $g^{'}(\hat{S}^{*})=0,$ we have $\hat{S}^{*}\log N\hat{S}^{*}=$
\[
\left[\frac{V}{MQ_{\max}}+\frac{\delta}{2K}\left(\log\left(P\log N\hat{S}^{*}\right)+\frac{1}{\log N\hat{S}^{*}}-\frac{1}{\hat{S}^{*}\log N\hat{S}^{*}}\right)\right]^{-1}.
\]

Therefore, we have $N\hat{S}^{*}\log N\hat{S}^{*}\leq\left(\frac{V}{MQ_{\max}}\right)^{-1}N=\frac{MNQ_{\max}}{V}\triangleq c_{1}$,
for $\hat{S}^{*}\geq3$ and all $\delta\geq0$. Thus we have $\hat{S}^{*}\leq\frac{1}{N}e^{W(c_{1})}$.
Note that, under $\delta\to0$, we have $\hat{Q}_{\pi(k)}\downarrow Q_{\pi(k)}$
and $\frac{\delta}{2K}\left(\log\left(P\log NK\right)+\frac{1}{\log NK}-\frac{1}{\hat{S}^{*}\log N\hat{S}^{*}}\right)\to0$,
which means the upper bound is achieved when $Q_{\pi(k)}\approx Q_{\max}$.

Note that, in the outer subproblem (\ref{eq:outer-subproblem-2}),
increasing $Q_{\pi(k)}$ to $\hat{Q}_{\pi(k)}$ for every $k$ yields
a larger solution point $\hat{S}^{*}(Q_{\max})\geq S^{*}(\mathbf{Q})$
{[}due the term $\sum_{k=1}^{S}Q_{\pi(k)}${]}. Hence, we have $S^{*}(\mathbf{Q})\leq\hat{S}^{*}(Q_{\max})\leq\frac{1}{N}e^{W(c_{1})}$.

\section{Proof of Theorem \ref{thm:proposed-scheme-rate-function} }

\label{app:LDP-rate-proposed-scheme}

In Lemma \ref{lem:Property-D_max}, the departure rate $D_{\max,b}(t;S)$
can be approximately given in (\ref{eq:departure-rate-baseline}),
which is a decreasing function of $S$ and has a Poisson distribution
with mean $\overline{D}_{\max,b}(t;S)=\frac{M\hat{\eta}(S)}{L}$.
With Lemma \ref{lem:Property-D_max}-\ref{lem:solution-S}, we have
$D_{\max,p}(t;S^{*})\geq D_{\max,b}(t;S^{*})\geq D_{\max,b}(t;\hat{S}^{*}(Q_{\max}))$,
since $S^{*}\leq\hat{S}^{*}$. Moreover, using the extreme value theorem,
we have $\overline{D}_{\max,b}/\frac{M}{LS}\log\left(P\log NK\right)\to1$,
as $K\to\infty$ \cite{sharif2005capacity}, which implies $\overline{D}_{\max,b}(t;\hat{S}(Q_{\max}))$
\[
\to\frac{M}{L\hat{S}^{*}(Q_{\max})}\log(P\log N\hat{S}^{*}(Q_{\max}))\triangleq\hat{\mu}_{p}(Q_{\max}).
\]

Consider the performance lower bound driven by the packet arrival
process $A(t)$ and departure process $D_{\max,b}(t,\hat{S}^{*}(Q_{\max}))$,
which are both Poisson processes. The corresponding LMF is given by
\begin{equation}
\hat{g}(x,\theta)=\lambda(e^{\theta}-1)+\hat{\mu}_{p}(x)\left(e^{-\theta}-1\right)\label{eq:LMF-proposed-T=00003D1}
\end{equation}
where $x=Q_{\max}$. Using Theorem \ref{thm:LDP_rate} and solving
$\hat{g}(x,\theta)=0$, we obtain $e^{\theta}=1$ and $e^{\theta}=\frac{\hat{\mu}_{p}(x)}{\lambda}$.
One can verify that $e^{\theta}=1$ only yields a trivial solution
$\hat{I}^{*}=0$. We thus calculate the lower bound rate function
by $\hat{I}^{*}=\int_{0}^{1}\log\frac{\hat{\mu}_{p}(x)}{\lambda}dx$.

Here, additional tricks should be used to complete the integral. Note
that when $Q_{\max}$ is small, $\hat{S}^{*}(Q_{\max})$ is small,
which violates the large $S$ assymptotic assumption to obtain the
approximated departure rate $D_{\max,b}(t,\hat{S}^{*}(Q_{\max}))$.
To fix this, we use the following augmented approximation, $\widetilde{\mu}_{p}(Q_{\max})=\max\left\{ \hat{\mu}_{p}(Q_{\max}),\frac{Mr_{0}}{LK}\right\} $,
where $r_{0}=\int_{0}^{\infty}\log(1+x)dF(x)$. Note that $r_{0}$
is the average per-beam data rate, and hence $\frac{Mr_{0}}{LK}$
is a lower bound average package departure rate for the maximum queue
process $Q_{\max}(t)$.

Note that $\hat{\mu}_{p}(x)$ is monotonically increasing. Define
$\epsilon_{K}$ as the solution to $\hat{\mu}_{p}(x)=\frac{Mr_{0}}{LK}$,
and $\epsilon=\inf\left\{ \epsilon_{K}:K\geq K_{0}\right\} $ for
some $K_{0}<\infty$. Using Theorem \ref{thm:LDP_rate}, we have 
\begin{eqnarray*}
\hat{I}^{*} & \ge & \int_{0}^{1}\log\frac{\widetilde{\mu}_{p}(x)}{\lambda}dx\\
 & = & \int_{0}^{1}\log\bigg(\frac{1}{\lambda_{tot}/K}\max\bigg\{\frac{M\log\left(P\log N\hat{S}^{*}(x)\right)}{L\hat{S}^{*}(x)},\\
 &  & \qquad\qquad\qquad\qquad\qquad\qquad\qquad\frac{Mr_{0}}{LK}\bigg\}\bigg)dx\\
 & = & \log\frac{M}{\lambda_{tot}L}+\int_{0}^{\epsilon}\log r_{0}dx\\
 &  & \qquad+\int_{\epsilon}^{1}\log\frac{\log\left(P\log N\hat{S}^{*}(x)\right)K}{\hat{S}^{*}(x)}dx\\
 & = & \log\frac{M}{\lambda_{tot}L}+\epsilon\log r_{0}+\left(1-\epsilon\right)\log K+C\triangleq I_{\mathrm{prop}}^{LB}
\end{eqnarray*}
where $C=\int_{\epsilon}^{1}\left\{ \log\left[N\log\left(PW\left(\frac{MNx}{V}\right)\right)\right]-W\left(\frac{MNx}{V}\right)\right\} dx$.
The first inequality is because $\widetilde{\mu}_{p}(Q_{\max})$ is
a lower bound estimation for the departure. 

Since $D_{\max,p}(t;S^{*})\geq D_{\max,b}(t;S^{*})$, we have $I_{prop}^{*}\geq\hat{I}^{*}$.
Thus we have proven the result.

\section{Proof of Theorem \ref{thm:rate-function-T-step}}

\label{app:proof-rate-function-T-step}

We first study the effect of the outdated QSI. Let $m(t)=\arg\max_{k}Q_{k}(t)$
be the user who has the longest queue at time $t$. Let $\mathcal{F}(t)$
deonte the feedback group under the proposed FFCA with $T=1$. We
concern with whether the feedback group $\mathcal{F}(t_{0})$ still
contains the longest queue user $m(t)$ at time $t$, i.e., the event
$m(t)\in\mathcal{F}(t_{0})$ happens at time $t$.

Consider the {}``\emph{best effort}'' event: the user $m(t_{0})$
is scheduled at every time slot but is still in the feedback group
$\mathcal{F}(t)$ at time $t$, 
\begin{eqnarray*}
 &  & \mathcal{E}_{BE}(t)\triangleq\bigg\{ Q_{\max}(t_{0})-\sum_{\tau=t_{0}}^{t}d_{m(t_{0})}(\tau)\\
 &  & \qquad+\sum_{\tau=t_{0}}^{t}A_{m(t_{0})}(\tau)>Q_{\pi^{-}(t_{0})}(t_{0})+\sum_{\tau=t_{0}}^{t}A_{\pi^{-}(t_{0})}(\tau)\bigg\}
\end{eqnarray*}
where $d_{m(t_{0})}(H_{m(t_{0})}(\tau))$ is the packet departure
rate under a \emph{fictitious }{}``\emph{best effort}'' policy that
schedules user $m(t_{0})$ at every time slot regardlessly of $\mathbf{Q}(\tau)$.
Specifically, according to (\ref{eq:snr_cdf}), the distribution of
$d$ is given by 
\begin{eqnarray*}
\mbox{Pr}(d\leq x) & = & \mbox{Pr}(\log(1+P\mbox{SINR})\leq x)\\
 & = & \mbox{Pr}(\mbox{SINR}\leq P^{-1}(2^{x}-1))\\
 & = & F(P^{-1}(2^{x}-1)).
\end{eqnarray*}
In addition, $\pi^{-}(t_{0})=\pi(S^{*}[\mathbf{Q}(t_{0})]+1)$ is
the user who just cannot be selected in the feedback set $\mathcal{F}(t_{0})$
at $t_{0}$. (Just recall that $\pi(\centerdot)$ is the ordered permutation
of $\mathbf{Q}$.) In $\mathcal{E}_{BE}$, one schedules the outdated
longest queue user $m(t_{0})$ at every time slot, but still, no user
from outside $\mathcal{F}(t_{0})$ has the longest queue at time $t$.
Note that we must have $Q_{m(t_{0})}(t)\geq Q_{m(t_{0})}^{BE}(t)$
almost surely, where $Q_{m(t_{0})}(t)$ is the queue length for user
$m(t_{0})$ under the queue-weighted scheduling in Stage II, and $Q_{m(t_{0})}^{BE}(t)$
is under the {}``\emph{best effort}'' scheduling. Therefore, we
must have $\mbox{Pr}\{m(t)\in\mathcal{F}(t_{0})\}\geq\mbox{Pr}\{\mathcal{E}_{BE}(t)\}$,
for $t_{0}\leq t\leq t_{0}+T-1$. The upper bound is tight in the
heavy queue region for small $T$.

Moreover, since $Q_{\max}(t_{0})>Q_{\pi^{-}(t_{0})}(t_{0})$, under
the i.i.d. assumption for the arrivals $A_{k}(t)$ and the CSI $H_{k}(t)$
respectively, we have 
\[
\mbox{Pr}(\mathcal{E}_{BE}(t))\geq\mbox{Pr}\big\{\sum_{\tau=t_{0}}^{t}v(\tau)>0\big\}\triangleq P_{0}^{t-t_{0}}\geq P_{0}^{T}
\]
where $v(\tau)=A_{1}(\tau)-A_{2}(\tau)-d(\tau)$. The last inequality
holds, since $\mathbb{E}v(\tau)<0$ and $\sum_{\tau=1}^{\delta}v(\tau)$
is more negative as $t-t_{0}$ increases.

We then study the departure rate for the process $Q_{\max}(t)$. Denote
$D_{\max}^{(T)}(\mathbf{H}(t),\mathbf{Q}(t);S^{*}(\mathbf{Q}(t_{0}),\mathcal{F}(t_{0}))$
as the packet departure for $Q_{\max}^{(T)}(t)$ under the $T$-step
feedback policy in $t_{0}\leq t\leq t_{0}+T-1$, where the feedback
probability is updated at time $t_{0}$. Similarly, denote $D_{\max}(\mathbf{H}(t),\mathbf{Q}(t);S^{*}(\mathbf{Q}(t),\mathcal{F}(t))$
as the packet departure under the per time slot feedback policy update
($T=1$). We have, \\$D_{\max}^{(T)}(\mathbf{H}(t),\mathbf{Q}(t);S^{*}(\mathbf{Q}(t_{0}),\mathcal{F}(t_{0}))$
\begin{eqnarray*}
 & \approx & D_{\max}(\mathbf{H}(t),\mathbf{Q}(t);S^{*}(\mathbf{Q}(t),\mathcal{F}(t))\centerdot1\{m(t)\in\mathcal{F}(t_{0})\}\\
 & \geq & D_{\max}(\mathbf{H}(t),\mathbf{Q}(t);S^{*}(\mathbf{Q}(t),\mathcal{F}(t))\centerdot1\{\mathcal{E}_{BE}(t)\}
\end{eqnarray*}
where the lower bound is tight in heavy queue region and $T$ is small.
The first approximate equality holds, since when the user with the
maximum queue is outside the feedback group under outdated QSI, $Q_{\max}(t)$
cannot be served at all. 

According to Theorem \ref{thm:LDP_rate}, We then need to find the
solution of the LMF $\widetilde{g}(x,\theta_{T}^{*})=0$ under the
$T$-step policy. The LMF of the random variable $D_{\max}^{(T)}\centerdot1\{\mathcal{E}_{BE}\}$
is given by 
\begin{eqnarray*}
\Lambda_{\widetilde{D}}^{T}(\theta) & \triangleq & \log\mathbb{E}[\exp(\theta D_{\max}(\centerdot)1\{\mathcal{E}_{BE}(t)\}]\\
 & = & \log\mathbb{E}\big\{\mathbb{E}[\exp(\theta D_{\max}(\centerdot)1\{\mathcal{E}_{BE}(t)\}]\big|1\{\mathcal{E}_{BE}(t)\}\big\}\\
 & = & \log\left(1-P_{0}^{T}+P_{0}^{T}\Lambda_{D}(\theta)\right)
\end{eqnarray*}
and the local LMF for the queuing process $Q_{\max}(t)$ is 
\[
\widetilde{g}(x,\theta^{(T)})=\Lambda_{A}(\theta)+\log\left(1-P_{0}^{T}+P_{0}^{T}\mathcal{M}_{D}(x,-\theta)\right)
\]
where $\mathcal{M}_{D}(x,-\theta)$ is the MGF of $D_{\max}(\centerdot)$.

To find the root $\theta_{T}^{*}(x)$ of the above function, we consider
a linearization, $\widetilde{g}_{L}(x,\theta_{T})=\widetilde{g}(x,\theta_{0}(x))+\nabla_{\theta}\widetilde{g}(x,\theta_{0}(x))\triangle\theta$,
where $\theta_{0}(x)$ is the solution to $\hat{g}(x,\theta(x))=0$
in (\ref{eq:LMF-proposed-T=00003D1}) under the $T=1$ policy. Let
$\beta_{0}\triangleq e^{\theta_{0}}$ and $\triangle\beta\approx e^{\theta_{T}}-\beta_{0}$.
Setting $\widetilde{g}_{L}(x,\theta_{T})=0$, we obtain, 
\begin{eqnarray*}
\frac{\triangle\beta(x)}{\beta_{0}(x)} & = & -\frac{\hat{\mu}_{p}(x)-\lambda+\log\left(1-P_{0}^{T}+P_{0}^{T}e^{(\lambda-\hat{\mu}_{p}(x))}\right)}{\mu(x)-\lambda\frac{P_{0}^{T}e^{(\lambda-\hat{\mu}_{p}(x))}}{1-P_{0}^{T}+P_{0}^{T}e^{(\lambda-\hat{\mu}_{p}(x))}}}\\
 & \geq & -\frac{\hat{\mu}_{p}(x)-\lambda+\log\left(1-P_{0}^{T}+P_{0}^{T}e^{(\lambda-\hat{\mu}_{p}(x))}\right)}{\mu(x)-\lambda}\\
 & = & -\frac{1}{\hat{\mu}_{p}(x)-\lambda}\log\left(e^{\hat{\mu}_{p}(x)-\lambda}-(e^{\hat{\mu}_{p}(x)-\lambda}-1)P_{0}^{T}\right)\\
 & \triangleq & \rho(x).
\end{eqnarray*}
The approximation, which is obtained by linearization, becomes accurate
when $P_{0}^{T}$ is close to $1$. Therefore, using Theorem \ref{thm:LDP_rate},
the rate function under the $T$-step feedback policy is bounded by
\begin{eqnarray*}
I_{\mathrm{prop}}^{(T)*} & \geq & \int_{0}^{1}\theta_{T}^{*}(x)dx=\int_{0}^{1}\log\beta_{0}\left(1+\frac{\triangle\beta(x)}{\beta_{0}(x)}\right)dx\\
 & \geq & I_{\mathrm{prop}}^{LB}-\int_{0}^{1}\rho(x)dx.
\end{eqnarray*}

\bibliographystyle{IEEEtran}
\bibliography{Delay_MIMO_bib,/Users/Allen/Dropbox/Draft/Bib_Classical_paper,Bib_Converge_Analysis}

\end{document}